# Methodologies for $^{176}$Lu-$^{176}$Hf Analysis of Zircon Grains from the Moon and Beyond

Xi Chen[1], Nicolas Dauphas[1*], Zhe J. Zhang[1], Blair Schoene[2], Melanie Barboni[3], Ingo Leya[4], Junjun Zhang[1], Dawid Szymanowski[2], Kevin D. McKeegan[5]

[1]Origins Laboratory, Department of the Geophysical Sciences and Enrico Fermi Institute, The University of Chicago, Chicago, IL 60637, USA

[2]Department of Geosciences, Princeton University, Princeton, NJ 08544, USA

[3]CLAS-NS Departments, Arizona State University, Tempe, AZ 85281, USA

[4]Physics Institute, University of Bern, Sidlerstrasse 5, 3012 Bern, Switzerland

[5]Department of Earth, Planetary, and Space Sciences, University of California, Los Angeles, CA 90095, USA

***To whom correspondence should be addressed** (dauphas@uchicago.edu)

**Keywords:** zircon, dating, extraterrestrial, cosmogenic, differentiation

**Abstract**

Zircons are found in extraterrestrial rocks from the Moon, Mars, and some differentiated meteorite parent-bodies. These zircons are rare, often of small size, and have been affected by neutron capture induced by cosmic ray exposure. The application of the $^{176}$Lu-$^{176}$Hf decay system to zircons from planetary bodies such as the Moon can help establish the chronology of large-scale differentiation processes, like the crystallization of the lunar magma ocean. Here, we present methods to measure the isotopic composition of Hf of extraterrestrial zircons dated using ID-TIMS U-Pb after chemical abrasion. We introduce a 2-stage elution scheme to separate Hf from Zr while preserving the unused Zr fraction for future isotopic analysis. The effect of neutron capture is also re-examined using the latest thermal neutron capture cross sections and epithermal resonance integrals. Our tests show that the precision of Hf isotopic analyses is close to what is theoretically attainable. We have tested this method to a limited set of zircon grains from lunar rocks returned by the Apollo missions (lunar soil 14163, fragmental polymict breccia 72275, and clast-rich breccia 14321). The model ages align with previously reported values, but further work is needed to assess the chronology of lunar magma ocean crystallization as only a handful of small zircons (5 zircons from 3 samples) were analyzed, and the precision of the analyses can be improved by measuring more and larger lunar zircon grains.

## INTRODUCTION

Zircon is a prime target mineral for $^{176}$Lu-$^{176}$Hf studies ($\lambda = 1.867 \times 10^{-11}$ yr$^{-1}$, $t_{1/2}$=37.12 Ga[1]) because it can readily be dated using U-Pb geochronology, has a low Lu/Hf ratio, and typically contains percent-level amounts of Hf. One can therefore measure present-day $^{176}$Hf/$^{177}$Hf ratios in single zircon grains, either in bulk or through *in situ* techniques, and calculate the initial $^{176}$Hf/$^{177}$Hf of the zircon with minimal correction for *in situ* decay of $^{176}$Lu. Initial $^{176}$Hf/$^{177}$Hf ratios can in turn be used to establish the history of planetary differentiation [2-10]. Zircons are relatively abundant in terrestrial rocks, and they are also found in lunar rocks, and in meteorites from Mars and Vesta. Studies of extraterrestrial zircons present specific challenges that are seldom encountered in terrestrial rocks, arising from the scarcity and scientific value of their host rocks, and the need to correct for shifts in Hf isotopic abundances induced by exposure to cosmic rays in space. D'Abzac *et al.*[11] and Bauer *et al.*[12] developed protocols for measuring the isotopic composition of Hf in small zircon and baddeleyite grains. They did not purify Hf and opted instead to monitor and correct isobaric interferences during analysis. Bast *et al.*[13] also focused on small samples, but they purified Hf before isotopic analysis using a two-stage ion-exchange chromatography for Lu-Hf dating. We also employ ion chromatography for Hf purification, but our method is tailored for extraterrestrial zircons, where factors like cosmogenic effects and normalization to chondrites are of concern.

Lunar zircons likely formed by either (*i*) crystallization of the lunar magma ocean[14, 15] from a liquid named KREEP that is highly enriched in incompatible elements such as K, REE, and P [16, 17] or (*ii*) later impact-induced melting[18, 19]. KREEP is found in dilute form in basalts and soils recovered from the Moon by the Apollo missions. The KREEP component is highly enriched in Zr relative to the bulk silicate Moon (by a factor of ~165[17]), leading to zircon crystallization in KREEP-rich magmas. Combining U-Pb and $^{176}Lu$-$^{176}Hf$ analyses of lunar zircons, one can estimate the initial $^{176}Hf/^{177}Hf$ ratio at the time of crystallization of lunar zircons, which should represent a snapshot of the composition of the KREEP reservoir at that time. KREEP is an enriched reservoir characterized by a low Lu/Hf ratio and unradiogenic $^{176}Hf/^{177}Hf$ ratios relative to the bulk silicate Moon (BSM) and the Chondritic Uniform Reservoir (CHUR). Comparison of the initial $^{176}Hf/^{177}Hf$ ratios of zircons with the inferred BSM value at the time of zircon crystallization can yield tight constraints on the time of LMO crystallization, which provides a minimum age for the formation of the Moon itself[2, 10].

Taylor *et al.*[10] analyzed the Hf isotopic compositions of lunar zircon grains from three polymict breccias and a soil collected by the Apollo 14 mission. The $^{176}Lu$-$^{176}Hf$ isotopic analyses were performed by laser-ablation multi-collector inductively coupled plasma mass spectrometry (LA-MC-ICP-MS) on zircons that had been dated using U-Pb geochronology by secondary ionization mass spectrometry (SIMS). A virtue of *in situ* isotopic analyses is that the samples are not totally consumed during analysis and such analyses can resolve complex zircon growth histories, which can confound bulk zircon analyses[7]. The main limitations of *in situ* analyses are precision and accuracy since some isobaric interferences such as $^{176}Lu$ and $^{176}Yb$ on $^{176}Hf$ cannot be eliminated and must be corrected for. The data of Taylor *et al.*[10] favored termination of lunar magma ocean (LMO) crystallization ~70 Myr (~4500 Ma) after solar system formation but were permissive of an age as late as 120 Myr (~4450 Ma). To better define that age, Barboni *et al.*[2] measured $^{176}Hf/^{177}Hf$ ratios by MC-ICP-MS after zircon digestion, while U-Pb dates were obtained by isotope dilution thermal ionization mass spectrometry (ID-TIMS) on chemically abraded zircons. The samples that they analyzed were four zircon fragments from the same sample set that had been studied previously by Taylor *et al.*[10], and an additional four affected by larger neutron capture effects. Barboni *et al.*[2] found very unradiogenic $^{176}Hf/^{177}Hf$ ratios in several zircons, suggesting early crystallization of the LMO. They concluded that LMO crystallization must have been completed within ~60 Myr (>4507 Ma) of the formation of the solar system, but the precision of the Hf isotopic measurements was limited and only a handful of samples were analyzed. A model age of KREEP was also estimated using $^{147}Sm$-$^{143}Nd$ and $^{176}Lu$-$^{176}Hf$ isochron analyses of KREEP-rich rock samples (Borg and Carlson[20] and references therein). These two decay systems yielded rock-scale model ages of 4334±37 Ma and 4356±37 Ma for KREEP (~220 Myr after solar system formation). There is thus considerable uncertainty on when LMO crystallization finished, with zircon and rock model ages giving values between ~60 and ~220 Myr after solar system formation[2, 10, 21-23]. Taylor *et al.*[10] and Barboni *et al.*[2] reported precisions on $^{176}Hf/^{177}Hf$ measurements of 1 to 4 ε-units at 2σ ($\varepsilon^{176}Hf$ is the deviation in parts per $10^4$ of the $^{176}Hf/^{177}Hf$ ratio relative to a reference material) on grain fragments that were 60-150 μm in size originally, but part of those grains had been consumed by prior laser ablation work. Higher precision and accuracy measurements are needed to provide more robust constraints on the formation of KREEP, which represents a firm minimum limit on the age of the Moon itself[2].

Combined $^{176}Hf/^{177}Hf$ and U-Pb measurements of zircons can also provide insights into the early differentiation history of Mars. Bouvier *et al.*[24] and Costa *et al.*[6] studied ancient zircons (4.43 to 4.48 Ga $^{207}Pb/^{206}Pb$ ages) extracted from martian polymict breccia NWA 7533/NWA 7034, also known as Black Beauty. The $^{176}Hf/^{177}Hf$ ratios measured in those zircons pointed to the existence of an enriched crustal reservoir on Mars that formed ~20 Myr after solar system formation. The extracted Martian zircon grains were 30 to 80 μm in size and the precision of $\varepsilon^{176}Hf$ isotopic analyses ranged from ~ ±0.3 to ±1.

Iizuka *et al.*[25] measured $^{176}Hf/^{177}Hf$ ratios in zircon grains extracted from the Agoult eucrite meteorite. Eucrites are basaltic meteorites that are thought to have formed in the crust of asteroid Vesta soon after the formation of the solar system. Iizuka *et al.*[25] used these measurements to constrain the solar system initial $^{176}Hf/^{177}Hf$ ratio (0.279781±0.000018). The typical size of zircon grains in eucrites is ~20 μm but those extracted from Agoult were ~80 μm. The precisions obtained on $\varepsilon^{176}Hf$ isotopic analyses in these zircons ranged from ~±0.3 to ±1.

Extraterrestrial zircons are precious and data quality is paramount as a handful of measurements can have large scale implications on the chronology of planetary differentiation [2, 6, 10, 24, 25]. As a part of an effort to better understand the magmatic differentiation and early bombardment history of the Moon, we have developed an analytical protocol to analyze Lu-Hf isotope systematics in small single zircon grains. Building on previous studies[25-27 28], we developed a protocol to first isolate Zr and Hf from interfering Yb and Lu elements and then further purify Zr from Hf. Peters *et al.*[29] found that inefficient removal of Zr from the Hf cut leads to unusual mass bias behavior and matrix-dependent effects on measured Hf isotopic ratios when using Jet-sampler and X-skimmer cones. By getting rid of Zr, we can therefore take advantage of the higher sensitivity of these cones without compromising the accuracy of Hf isotopic analyses. Another motivation for this second step is to allow future isotopic analyses of Zr on the same sample aliquots for the study of Zr mass-dependent isotopic fractionation[30-39], nucleosynthetic anomalies[40-45], and decay of extinct radionuclide $^{92}Nb$ into $^{92}Zr$[46-49]. Hafnium isotopic analyses are done by MC-ICP-MS. We also compare achieved and theoretically attainable precisions for internally normalized ratios based on counting statistics and Johnson noise.

**METHODS**

The method developed combines chemical abrasion of zircons, U-Pb dating by TIMS, Lu/Hf determination using a quadrupole ICP-MS, purification of Hf in a 2-stage chromatographic procedure using TODGA and Ln-Spec resin, and Hf isotopic analysis by MC-ICP-MS. The various steps involved are outlined in Fig. 1 and described in detail below.

**Zirconium and hafnium separation**

A two-stage procedure modified from Zhang[50] and Iizuka et al.[25] was developed for separating Zr and Hf from Yb, Lu and other interfering elements (**Table 1; Fig. 1**). In a first step, DGA (N,N,N',N'-tetra-n-octyldiglycolamide) resin from Eichrom (previously TODGA; now DGA normal[51, 52]) is used to collect a Zr-Hf cut, as described by Zhang *et al*.[53] in their protocol for Ti separation. In a second step, Ln-Spec resin is used to further separate heavy rare earth elements (REEs), Zr and Hf (**Table 1**)[50]).

The first step uses a 2-mL column of 0.8 cm diameter and 4 cm length, filled with TODGA resin. The resin in the column is cleaned using 10 mL of 3 M $HNO_3$, 10 mL of 3 M $HNO_3$+1vol% $H_2O_2$, and 4 mL $H_2O$. The resin is conditioned using 15 mL of 12 M $HNO_3$. The sample is then loaded in 10 mL of 12 M $HNO_3$. Loading the sample in 12 M $HNO_3$+1vol% $H_2O_2$ instead of 12 M $HNO_3$, as we have done, would expedite Ti elution and reduce slightly the blank. Titanium is eluted with 10 mL of 12 M $HNO_3$+1vol% $H_2O_2$. Iron is eluted with 10 mL of 3 M $HNO_3$. Finally, Zr and Hf are eluted together with 20 mL of 3 M $HNO_3$-0.3 M HF. The Zr and Hf cut is dried down on a hot plate and taken up in 0.5 mL of 2.5 M HCl.

To optimize the chemical separation procedure in the second step, the elution curve was calibrated using a multi-element standard solution containing Zr, Hf, and 24 other elements, including all the HFSEs and REEs (**Fig. 2**). Single element ICP-MS standard solutions (Spex CertiPrep) at concentrations of 1000 μg/mL were used to prepare this standard mixture. A similar calibration was done using solutions retrieved after U-Pb chemistry of terrestrial zircon reference materials (AS3, 91500) and synthetic zircon doped with REEs (MUNZirc 32a[54]) (**Fig. 3**). Zirconium and hafnium purifications of the standard mixture and the reference zircon solutions were done using a 0.35 mL fluoropolymer column (length = 20 cm, diameter = 1.5 mm) loaded with Ln-Spec resin (100 – 150 µm; HDEHP). Liquid was forced through the column using a pressure differential established by a vacuum box positioned below the column, with the vacuum pressure adjusted to maintain a flow rate of ~1-2 mL/hr. The resin was cleaned with 18 mL of 6 M HCl-0.06 M HF followed by 14 mL of 6 M HCl-0.2 M HF to ensure the removal of Zr and Hf that might be in the resin before loading the samples. The column filled with resin was preconditioned with 6 mL of 2.5 M HCl. The sample solutions were loaded onto the column in ~ 0.5 mL of 2.5 M HCl and matrix elements were removed with 12 mL of 6 M HCl-1vol% $H_2O_2$. Zirconium was first eluted in 22 mL of 6 M HCl-0.06 M HF. Hafnium was finally eluted in 7 mL of 6 M HCl-0.2 M HF (**Table 1**). Several studies have previously used Ln-spec resin for separating Zr, Lu, and Hf. Münker *et al.*[27] developed an elution protocol that can handle various rock types. They loaded the samples in 3 M HCl+0.1 M ascorbic acid, rinsed the matrix in 3M HCl, eluted Lu in 6 M HCl, eluted Ti in 0.09 M citric acid-0.4 M $HNO_3$-1vol% $H_2O_2$, eluted Zr in 6 M HCl-0.06 M HF, and eluted Hf in 6 M HCl-0.2 M HF. The acids used for eluting Zr and Hf are identical to ours. Iizuka *et al.*[25] used a simpler procedure for extraterrestrial zircons. They loaded their samples in 2.5 M HCl, removed the matrix in 6 M HCl+0.06 M HF, eluted Zr in 6 M HCl+0.06 M HF, and eluted Hf in 2 M HF. Our procedure uses 6 M HCl throughout, with other reagents added to target specific elements (1 vol% $H_2O_2$ for Ti, 0.06 M HF for Zr, 0.2 M HF for Hf), which simplifies reagent preparation.

Following chromatographic separation, the Zr and Hf cuts were dried down, taken back up in ~1 mL of aqua regia (3:1 mixture of $HCl:HNO_3$), and dried again before being re-dissolved in concentrated $HNO_3$. The re-dissolved solutions were dried again to near dryness (right before complete evaporation) and taken back up in 0.3 M $HNO_3$-0.07 M HF for Hf isotopic analysis. The Hf procedural blank was below detection limit (<5 pg) and was negligible compared to the amount of Hf in the single zircon grains analyzed (>4000 pg).

**Hafnium Isotope Mass spectrometry**

The Hf isotopic analyses were done on a Neptune MC-ICP-MS upgraded with a Pfeiffer OnTool Booster turbo pump to Neptune Plus specifications. The samples in 0.3 M $HNO_3$-0.07 M HF were injected into the Ar plasma torch using an Aridus II desolvating nebulizer. The sample, auxiliary, and cooling gas flows were set to ~0.825, 1, and 16 mL/min, respectively. The Ar and $N_2$ gas flows of the Aridus II nebulizer were set to 7.3 L/min and 0.14 mL/min, respectively. High-transmission Jet sample and X-skimmer cones were used. All measurements were done in low resolution. The purified Hf fractions were dissolved in 0.3 M $HNO_3$-0.07 M HF. The sensitivity was ~0.2 V/ppb on $^{177}$Hf (18.60%) measured on a $10^{11}$ Ω amplifier at a sample uptake rate of ∼100 μL/min. Isotopes $^{174}$Hf, $^{176}$Hf, $^{177}$Hf, $^{178}$Hf, $^{179}$Hf, and $^{180}$Hf as well as $^{172}$Yb, $^{175}$Lu, and $^{184}$W were measured in static mode on 9 Faraday cups with $^{172}$Yb and $^{174}$Hf Faraday cups connected to $10^{12}$ Ω amplifiers and other cups connected to $10^{11}$ Ω amplifiers. Isotope $^{172}$Yb was measured to monitor possible interferences from $^{174}$Yb and $^{176}$Yb. Isotope $^{175}$Lu was measured to monitor a possible interference from $^{176}$Lu. Isotope $^{184}$W was measured to monitor a possible interference from $^{180}$W. All potential isobaric interferences from Yb, Lu, and W on Hf isotopes were corrected for but these were always negligible, which is expected given the low Yb/Hf, Lu/Hf, and W/Hf ratios of zircons and the high selectivity of the Hf-Zr purification procedure outlined above. Hafnium was diluted to ~1-10 ppb for isotopic analysis. The measurements were divided into 30 cycles of 8.389 s integration time each. The 0.3 M $HNO_3$-0.07 M HF dilution medium was measured at the beginning and at the end of each sequence run and average intensities were subtracted from sample and standard measurements (on peak zero). Individual sample measurements were bracketed by the analysis of JMC-Hf 475 standard solutions whose concentrations were adjusted to match those of the samples that they bracket.

Natural processes[55] and mass spectrometry[56] induce Hf isotopic fractionation that must be corrected for before discussing $^{176}$Hf variations arising from decay of $^{176}$Lu. Mass bias ($\beta$) was calculated by normalizing $^{179}$Hf/$^{177}$Hf ratios in the samples and bracketing standards to a fixed reference value of 0.7325[57] using the exponential mass fractionation law $r_{2/1}=R_{2/1}(m_2/m_1)^b$ with $r_{2/1}$ and $R_{2/1}$ the measured (meas) and "unfractionated" (ref) ratios, respectively, and $m_2/m_1$ the ratio of the atomic masses of the two isotopes[58],

$$\beta = \ln\left[\left(\frac{^{179}\mathrm{Hf}}{^{177}\mathrm{Hf}}\right)_{\mathrm{meas}} \Big/ \left(\frac{^{179}\mathrm{Hf}}{^{177}\mathrm{Hf}}\right)_{\mathrm{ref}}\right] \Big/ \ln\left(m_{^{179}\mathrm{Hf}}/m_{^{177}\mathrm{Hf}}\right), \tag{1}$$

where $m_i$ is the atomic mass of isotope $i$ and $\left(^{179}\mathrm{Hf}/^{177}\mathrm{Hf}\right)_{\mathrm{ref}} = 0.7325$. The possible contributions of isobaric interferences on $^{176}$Hf and $^{180}$Hf were calculated assuming that Yb, Lu, and W would show approximately the same mass bias as Hf ($\beta_{\mathrm{Yb}} = \beta_{\mathrm{Lu}} = \beta_{\mathrm{W}} = \beta_{\mathrm{Hf}}$),

$$^{176}\mathrm{Hf} = I_{176} - I_{172}\left(\frac{^{176}\mathrm{Yb}}{^{172}\mathrm{Yb}}\right)_{\mathrm{ref}}\left(\frac{m_{^{176}\mathrm{Yb}}}{m_{^{172}\mathrm{Yb}}}\right)^{\beta} - I_{175}\left(\frac{^{176}\mathrm{Lu}}{^{175}\mathrm{Lu}}\right)_{\mathrm{ref}}\left(\frac{m_{^{176}\mathrm{Lu}}}{m_{^{175}\mathrm{Lu}}}\right)^{\beta}, \tag{2}$$

$$^{180}\mathrm{Hf} = I_{180} - I_{184}\left(\frac{^{180}\mathrm{W}}{^{184}\mathrm{W}}\right)_{\mathrm{ref}}\left(\frac{m_{^{180}\mathrm{W}}}{m_{^{184}\mathrm{W}}}\right)^{\beta}, \tag{3}$$

where $I_k$ is the ion intensity measured at mass $k$ after on-peak-zero subtraction, and "ref" stands for reference and corresponds to the typical terrestrial isotopic composition for the elements considered: $\left({}^{174}\mathrm{Yb}/{}^{172}\mathrm{Yb}\right)_{\mathrm{ref}} = 1.458085$, $\left({}^{176}\mathrm{Yb}/{}^{172}\mathrm{Yb}\right)_{\mathrm{ref}} = 0.584517$, $\left({}^{176}\mathrm{Lu}/{}^{175}\mathrm{Lu}\right)_{\mathrm{ref}} = 0.026525$, and $\left({}^{180}\mathrm{W}/{}^{184}\mathrm{W}\right)_{\mathrm{ref}} = 0.004086$. These interference corrections were always negligible but were implemented to streamline the data reduction procedure in case an outlier sample requiring significant correction was analyzed. After on-peak-zero (baseline) subtraction and correction of isobaric interferences, ${}^{i}\mathrm{Hf}/{}^{177}\mathrm{Hf}$ ratios are corrected for mass fractionation based on the mass bias calculated from the ${}^{179}\mathrm{Hf}/{}^{177}\mathrm{Hf}$ ratio (the star superscript indicates that the ratio has been corrected for mass-fractionation by internal normalization),

$$\left(\frac{{}^{i}\mathrm{Hf}}{{}^{177}\mathrm{Hf}}\right)^{*} = \left(\frac{{}^{i}\mathrm{Hf}}{{}^{177}\mathrm{Hf}}\right)_{\mathrm{meas}} \Big/ \left(\frac{m_{{}^{i}\mathrm{Hf}}}{m_{{}^{177}\mathrm{Hf}}}\right)^{\beta}, \quad (4)$$

The sample solutions were measured in a sequence standard-sample-standard (STD1-SMP-STD2), with the bracketing standard being a solution of JMC-Hf 475 diluted to a sample-matched concentration in the same acid mixture as the sample. For each bracket $j$, $\varepsilon^{i}\mathrm{Hf}$ values are calculated as,

$$\varepsilon^{i}\mathrm{Hf}_{\mathrm{SMP},j} = \left[\frac{2\left({}^{i}\mathrm{Hf}/{}^{177}\mathrm{Hf}\right)^{*}_{\mathrm{SMP},j}}{\left({}^{i}\mathrm{Hf}/{}^{177}\mathrm{Hf}\right)^{*}_{\mathrm{STD1},j}+\left({}^{i}\mathrm{Hf}/{}^{177}\mathrm{Hf}\right)^{*}_{\mathrm{STD2},j}} - 1\right] \times 10^{4}, \quad (5)$$

Depending on the zircon size and amount of Hf available, several bracket measurements were done ($n$=1 to 4; each bracket analysis consumed ~5 ng of Hf). The reported Hf isotopic composition for a sample is the average of these $n$ bracket measurements, $\varepsilon^{i}\mathrm{Hf}_{\mathrm{SMP}} = \sum_{j=1}^{n} \varepsilon^{i}\mathrm{Hf}_{\mathrm{SMP},j}/n$. For the purpose of comparing our data with previous studies, εHf values were also converted to absolute ratios using the following values for the isotopic ratios of the Johnson Matthey Company standard (JMC)-Hf 475: ${}^{180}\mathrm{Hf}/{}^{177}\mathrm{Hf} = 1.886666$, ${}^{178}\mathrm{Hf}/{}^{177}\mathrm{Hf} = 1.467168$, and ${}^{176}\mathrm{Hf}/{}^{177}\mathrm{Hf} = 0.282160$[59]. We also calculated $\varepsilon^{i}\mathrm{Hf}_{\mathrm{STD},p}$ of the standard bracketed by itself ($\varepsilon^{i}\mathrm{Hf}_{\mathrm{STD},p}$ is defined as the $\varepsilon^{i}\mathrm{Hf}$ value of $\mathrm{STD}_{p}$ bracketed by $\mathrm{STD}_{p-1}$ and $\mathrm{STD}_{p+1}$ in a sequence $STD_{p-1}$-SMP-$STD_p$-SMP-$STD_{p+1}$,). These standards were measured at the same concentration in the same condition as the sample, so we use the standard deviation of these standard bracket isotopic analyses $\sigma\left(\varepsilon^i\mathrm{Hf}_{\mathrm{STD}}\right)$ to calculate the uncertainty of the mean sample isotopic composition, $\sigma\left(\varepsilon^i\mathrm{Hf}_{\mathrm{SMP}}\right) = \sigma\left(\varepsilon^i\mathrm{Hf}_{\mathrm{STD}}\right)$. The reason for doing so is that there were more repeats of standard bracketed by standards than sample bracketed by standards, so calculation of the standard deviation is more reliable. Another approach to calculating uncertainties would be to use the dispersion of repeat cycles within a single analysis. We find good agreement between the two approaches (Fig. 4) but use repeat standard analyses to calculate error bars because it measures dispersion over a timespan that is more relevant to sample analyses (fluctuations happening in a matter of hours as opposed to minutes) and should be more reliable. All uncertainties are reported as 95% confidence interval (95% CI; 2σ).

## RESULTS AND DISCUSSION

### Limits on Precision of Hf isotopic analyses

We discuss below the precision that we were able to achieve and compare the results with what is theoretically possible. The floor to attainable precision by MC-ICP-MS equipped with traditional resistor-based signal amplification is set by counting statistics and Johnson (thermal) noise. Temperature, amplifier gain, and total number of ions captured in the Faraday cups will all influence the theoretically attainable precision. Dauphas *et al.*[60] calculated this theoretical limit for internally normalized ratios and the formula for $^{176}\mathrm{Hf}/^{177}\mathrm{Hf}$ ratio internally normalized using $^{179}\mathrm{Hf}/^{177}\mathrm{Hf}$ ratio is,

$$\sigma(\varepsilon^{176}\mathrm{Hf})^2 = 10^8\left[X_{176} + \left(\frac{\mu_{176/177}}{\mu_{179/177}}\right)^2 X_{179} + \left(\frac{\mu_{176/179}}{\mu_{179/177}}\right)^2 X_{177}\right], \quad (6)$$

where

$$X_i = \frac{eR_i}{U_i t} + \frac{4k_B T R_i}{{U_i}^2 t} , \tag{7}$$

$$\mu_{j/i} = \ln(j/i) , \tag{8}$$

with $e = 1.602 \times 10^{-19}\ C$ the elementary charge, $t$ (s) the duration of data acquisition, $k_B$ ($m^2 kg \cdot s^{-2} K^{-1}$) the Boltzmann constant, $T$ (K) the temperature of the feedback resistor, $R_i(\Omega)$ is the feedback-resistance of the amplifier used to measure mass $i$, and $U_i$ is the average voltage measured for mass $i$. The total acquisition time for a single bracket analysis (30 cycles) was 251.67 s. The sensitivity of the instrument for Hf was 1.2 V/ppb (the voltage corresponds to the sum of Hf isotopes), which we can use to calculate the voltages for all isotopes for a given Hf concentration in solution. This can also be converted to an amount of Hf consumed by multiplying the Hf solution concentration by the time and the nebulizer flow rate of 100 μL/min. The amplifiers are maintained at a temperature of 40 °C and $^{176}$Hf, $^{177}$Hf, $^{178}$Hf, $^{179}$Hf, $^{180}$Hf isotopes were measured using $10^{11}\ \Omega$ resistors. We used Eq. 8 to calculate the theoretical limit on precision and compared it with the standard deviation calculated from the multiple standard brackets (external precision hereafter). We also compared it with the internal precision calculated by taking the standard deviation of the mean (therefore SE) for all cycles acquired during a single analysis.

**Figure 4** shows the theoretically attainable precision curve (solid blue line) on $\varepsilon^{176}$Hf together with measured internal and external precisions. As expected, the uncertainty increases with lower Hf concentration. The measured internal and external uncertainties agree well with the theoretically achievable precisions, which demonstrates that the uncertainties are only limited by counting statistics and detector noise, and the amount of Hf available. At the beginning of each analytical session involving the analysis of precious extraterrestrial samples, we used the first day of measurements to test wheher the precision achieved was close to the theoretical limit.

## Accuracy of Hf isotopic analyses

To test the accuracy of the complete analytical procedure, we performed multiple analyses of three dissolved natural zircon reference materials (AS3, 91500) and one dissolved synthetic zircon doped with REE (MUNZirc 32a) (Table 2). All measurements were done with relatively low amounts of Hf, which did not exceed 20 ng, corresponding to a ~94 μm equivalent zircon grain diameter (calculated assuming ~1 wt% Hf in zircon). **Figure 5** shows the difference (expressed in $\Delta\varepsilon^{176}\text{Hf} = \left[\frac{(^{176}\text{Hf}/^{177}\text{Hf})_{\text{measured}}}{(^{176}\text{Hf}/^{177}\text{Hf})_{\text{Literature}}} - 1\right] \times 10^4$ ) of these three zircons relative to recommended literature $^{176}Hf/^{177}Hf$ values of 0.282184±0.000016 for FC-1 (zircon from the same geological unit of AS3), 0.282306±0.000008 for 91500, and 0.282140±0.000005 for MUNZirc 32a (2SD)[54, 61]. The different data points are replicate analyses involving the whole purification procedure, starting from the same solution but processing different amounts of Hf through chromatography. All data points of 91500, and MUNZirc 32a fall within the error of their expected literature values, meaning that our methodology yields accurate analyses down to 3 ng total Hf consumed. The $^{176}Hf/^{177}Hf$ value that we measure for AS3 is larger than the reference value by ~+0.5 ε[61], independently of sample size ($^{177}Hf$ signal). This small, but statistically significant discrepancy is unlikely due to unresolved mass interferences because the analyses of MUNZirc 32a are accurate despite the much higher concentrations of potentially interfering elements in that reference material[54]. The difference in $^{176}Hf/^{177}Hf$ ratio is more likely due to heterogeneity between the FC-1 sample measured by Woodhead and Hergt [61] and our AS3 zircons.

Having established that the measurements are accurate, and that precision follows the theoretical limit imposed by counting statistics and Johnson noise, we can estimate the precision attainable for samples from different planetary bodies. The developed method would allow us to measure $\varepsilon^{176}Hf$ in typical extraterrestrial zircons with precisions of ±1 ε in ~50 μm-size zircons.

**Extraterrestrial test materials**

Interpretation of measured $\varepsilon^{176}Hf$ in zircons that have been independently dated using U-Pb requires consideration of *in situ* production of $^{176}Hf$ from $^{176}Lu$ decay and comparison with CHUR. Knowing the initial $\varepsilon^{176}Hf$ values and age of the zircons, it is possible to calculate model ages of crustal differentiation, as enriched reservoirs are characterized by low Lu/Hf ratio and thus unradiogenic $\varepsilon^{176}Hf$ values[57]. With extraterrestrial samples, additional complications arise from the fact that isotopic ratios can be modified by interaction with cosmic rays and nucleosynthetic anomalies may be present.

Nucleosynthetic anomalies reflect the fact that the solar system was never fully homogenized and different planetary bodies received different proportions of products of stellar nucleosynthesis[62, 63]. The search for nucleosynthetic anomalies in planetary materials has focused on isotopes other than $^{176}Hf$ because decay of $^{176}Lu$ obscures potential nucleosynthetic effects on $^{176}Hf$[42, 64, 65]. Sprung et al.[66] evaluated the effect of nucleosynthetic heterogeneities on the $^{176}Lu$-$^{176}Hf$ system. They measured several bulk meteorites and the only isotopic variations that they found were cosmogenic in nature, with no evidence for a distinct nucleosynthetic contribution. They did find hints for the presence of nucleosynthetic anomalies in meteoritic refractory inclusions, corresponding to possible inherited $\varepsilon^{176}Hf$ variations of up to ~1.5 ε. More meteorite Hf isotope measurements have been performed since the Sprung *et al.*[66] study that have revealed isotopic anomalies in acid leachates[42, 64, 65] and refractory inclusions[41], but no anomaly in bulk rocks[41, 66]. Cosmogenic and nucleosynthetic effects form almost orthogonal trends in $\varepsilon^{180}Hf$-$\varepsilon^{178}Hf$ space (**Fig. 6**, Sprung et al.[66]), so we can combine those measurements with the known slopes imparted by cosmogenic and nucleosynthetic[42, 64, 65] effects to estimate the range of allowable nucleosynthetic $\varepsilon^{180}$Hf and $\varepsilon^{178}$Hf variations in meteorites (subscript *n* stands for original nucleosynthetic signature corrected for cosmogenic effects),

$$\varepsilon^{178}\mathrm{Hf}_n = \frac{\varepsilon^{180}\mathrm{Hf}_m - s_{c,180/178}\varepsilon^{178}\mathrm{Hf}_m}{s_{n,180/178} - s_{c,180/178}}, \quad (9)$$

$$\varepsilon^{180}\mathrm{Hf}_n = s_{n,180/178}\varepsilon^{178}\mathrm{Hf}_n. \quad (10)$$

where $\varepsilon^{178}\mathrm{Hf}_m$ and $\varepsilon^{180}\mathrm{Hf}_m$ are the measured compositions, $s_{n,180/178} = 1.215$ and $s_{c,180/178} = -1.53$ are the slopes between $\varepsilon^{180}$Hf and $\varepsilon^{178}$Hf imparted by nucleosynthetic and cosmogenic effects, respectively. We can then use those compositions to estimate the possible nucleosynthetic shift in $^{176}$Hf,

$$\Delta\varepsilon^{176}\mathrm{Hf}_n = s_{n,176/178}\varepsilon^{178}\mathrm{Hf}_n, \quad (11)$$

$$\Delta\varepsilon^{176}\mathrm{Hf}_n = s_{n,176/180}\varepsilon^{180}\mathrm{Hf}_n, \quad (12)$$

with $s_{n,176/178} = 3.13$ and $s_{n,176/180} = 2.57$, as derived from *s*-process calculations using the formula of Dauphas *et al.*[60, 67] and data from Bisterzo *et al.*[68] . In Fig. 6, we plot the expected relationship between nucleosynthetic anomalies (Eqs. 12, 13 and 14), together with measurement-derived nucleosynthetic shifts in $\varepsilon^{178}$Hf and $\varepsilon^{180}$Hf. As shown, $\varepsilon^{178}$Hf gives the tightest constraint on the nucleosynthetic contribution on $\varepsilon^{176}$Hf and given the lack of isotopic anomalies for meteorites at a bulk scale, we expect effect of nucleosynthetic anomalies on $\varepsilon^{176}$Hf to be significantly less than ±0.3 ε, which is small compared to isotopic variations arising from $^{176}$Lu decay. The Moon is known to have very similar isotopic composition to Earth (Dauphas and Schauble[63] and references therein), so the Moon should have started with a Hf isotopic composition almost identical to the terrestrial composition. Therefore, we can neglect any potential isotopic shift due to inherited anomalies.

A more important consideration in $^{176}$Lu-$^{176}$Hf analyses of extraterrestrial samples is the presence of cosmogenic effects produced by cosmic ray exposure both at the surface of the object or during transit to Earth in the case of meteorites. Hafnium isotopes in terrestrial samples are not significantly affected by cosmic rays because Earth's surface is partly shielded by the atmosphere and magnetosphere and rocks at Earth's surface are constantly eroded. To illustrate how cosmogenic effects can be tackled, we have studied 5 lunar zircons. The zircons that we targeted are small and the measurements have relatively low precision. The zircon grains were hand-picked from samples recovered by the Apollo missions: lunar soil 14163, fragmental polymict breccia 72275, and clast-rich breccia 14321. The zircons range in size from 150-300 μm, which is on the lower end of the size distribution for lunar zircons studied thus far. They were chemically abraded using the technique outlined in Barboni *et al.*[2] and references therein (Fig. 1). Zircon fragments were removed from epoxy mounts and thermally annealed by transferring the fragments into quartz crucibles and heating to 900 °C for 48 hours. Fragments were then rinsed with acetone in 3-mL fluoropolymer PFA beakers, leached in 6 M HCl at 100 °C for one hour, and rinsed again using milliQ $H_2O$. The zircon fragments were then loaded into 200 μL Savillex "micro"-capsules with 100 μL 29 M HF + 15 μL 3 M $HNO_3$ for a first step of chemical abrasion[69] in Parr bombs at 185 °C for 6 hours. Grains were rinsed 10 times after the 6-hour step-1 leaching with 6 M HCl, milliQ $H_2O$ and 29 M HF before being loaded again into micro-capsules with 100 μL 29 M HF + 15 μL 3 M $HNO_3$ for a second 6-hour step of leaching at 185°C. All the rinses from each zircon were collected in a separate Teflon beaker as leachate L1. The same rinsing procedure as was done for the step-1 leaching was also applied after step-2 leaching (saved as L2 for each zircon). The remaining zircon fragments (subsequently referred to as "residues") were then spiked and dissolved to completion in 100 μL 29 M HF + 15 μL 3 M $HNO_3$ in Parr bombs at 210°C for 48 hours. Leachate L1, L2 and zircon fragment residues were spiked with the EARTHTIME $^{202}$Pb–$^{205}$Pb–$^{233}$U–$^{235}$U tracer and allowed to equilibrate either on a hotplate (L1 and L2) or during dissolution (residues)[70, 71]. The dissolved residue and leachates L1 and L2 were individually dried down and converted to chlorides by overnight re-dissolution using 200 µL of 6N HCl on a hotplate. They were subsequently dried down and brought up in 50 µL 3 M HCl. Those dry-down/redissolution steps ensure complete sample-spike equilibration. The U-Pb and trace element (including Lu and Hf) fractions were separated by anion exchange column chromatography using a single 50 µL column and AG-1 X8 resin (200–400 mesh; Cl-form) from Eichrom[72]. This procedure involves elution of Zr, Hf and other trace elements in ~200 µl 3N HCl, which was aliquoted in equal parts and saved for Hf isotopic analysis by MC-ICP-MS and trace element (including Lu/Hf ratio) analysis by single-collector ICP-MS. The U–Pb elutions were collected in single beakers, dried down with a drop of 0.02 M $H_3PO_4$, and were analyzed on a single outgassed zone-refined Re filament in Si-gel emitter[73] using an Isotopx Phoenix TIMS at Princeton University. For most zircons, we analyzed several leachate fractions and the residue, so the five zircons yielded nine ages and nine $\varepsilon^{176}$Hf measurements (**Table 3 and Table S3**).

**Correction for radiogenic ingrowth, neutron capture effects, and normalization to CHUR**

In each zircon, $\varepsilon^{176}\text{Hf}_{\text{CHUR}}$ is first corrected for neutron capture (NC) effects[2, 21, 22, 66], which mainly change the $^{179}$Hf/$^{177}$Hf ratio used for mass bias correction through $^{177}$Hf(*n,g*)$^{178}$Hf and $^{178}$Hf(*n,g*)$^{179}$Hf reactions. Neutron capture effects on $^{176}$Hf/$^{177}$Hf (including those arising indirectly from the mass bias correction) are corrected for by monitoring variations in internally normalized $^{178}$Hf/$^{177}$Hf and $^{180}$Hf/$^{177}$Hf ratios [2, 21, 22, 66]. Our correction procedure builds on the work of Sprung *et al.*[21, 66], who studied neutron capture effects on Hf and Sm isotopes in lunar samples. The different Hf isotopes are affected differently by thermal (<0.5 eV) and epithermal (>0.5 eV) secondary neutrons because of isotope specific neutron capture cross sections and resonance integrals. The neutron energy distribution can vary depending on target composition, and to a lesser extent depth and irradiation geometry. The $\varepsilon^{180}$Hf/$\varepsilon^{149}$Sm ratio provides a sensitive measure of the ratio of epithermal-to-thermal neutron fluences (ep/th) because $\varepsilon^{180}$Hf is mostly affected by epithermal neutrons while $\varepsilon^{149}$Sm is mostly affected by thermal neutrons. Sprung *et al.*[21] adjusted the epithermal and thermal neutron capture fluences in each sample to reproduce the measured $\varepsilon^{180}$Hf/$\varepsilon^{149}$Sm ratio. Because thermal neutron capture cross sections and resonance integrals are relatively well determined for Hf isotopes, with knowledge of the ep/th neutron fluence, it is possible to calculate the corresponding cosmogenic shift in $\varepsilon^{176}$Hf. Measuring Sm or Gd isotope ratios in small single lunar zircon grains is difficult if not impossible (a lunar zircon would typically contain only ~4 pg of Sm and ~16 pg of Gd), so Barboni *et al.*[2] used the correlation between calculated cosmogenic $\varepsilon^{176}$Hf and measured $\varepsilon^{178}$Hf shifts to correct their data for neutron capture effects. Using combined Sm and Hf isotopic analyses of low- and high-Ti lunar basalts and KREEP-rich samples, Sprung *et al.*[21] showed that there is a tight correlation between shifts in $\varepsilon^{180}$Hf and $\varepsilon^{178}$Hf and that this correlation is insensitive to varying ep/th neutron exposure spanning a range from 0.44 to 2.2 (determined from $\varepsilon^{149}$Sm). This suggests that cosmogenic shifts in Hf isotopic compositions are not very dependent on the neutron energy distribution (ep/th ratio) and that either $\varepsilon^{180}$Hf or $\varepsilon^{178}$Hf can be monitored to quantify the cosmogenic correction on $\varepsilon^{176}$Hf arising either directly from *n*-capture on $^{176}$Hf and $^{177}$Hf or indirectly via the mass bias correction (*i.e.*, $^{179}$Hf/$^{177}$Hf).

Because cosmogenic shifts on $^{176}$Hf/$^{177}$Hf can amount to several ε units, we have re-evaluated the effect of variations in the energy spectrum of secondary neutrons to compare with the model results of Sprung *et al.*[21]. We consider thermal cross sections and resonance integrals taken from the Jeff-3.1 database[74, 75] and test whether our results depend on the choice of the nuclear database by comparing it with cross-sections and resonance integrals from the ENDFB-VIII database[76]. It turns out that the calculated cosmogenic shifts are essentially indistinguishable. Our calculations confirm that calculated $\Delta\varepsilon^{176}$Hf/ $\Delta\varepsilon^{178}$Hf and $\Delta\varepsilon^{176}$Hf/ $\Delta\varepsilon^{180}$Hf vary little with the ratio ep/th (**Fig. 7**). For pure thermal neutrons (ep/th=0), we have $\Delta\varepsilon^{176}$Hf/$\Delta\varepsilon^{178}$Hf=2.063 and $\Delta\varepsilon^{176}$Hf/$\Delta\varepsilon^{180}$Hf=-1.687. For pure epithermal neutrons (ep/th= $+\infty$ ), we have $\Delta\varepsilon^{176}$Hf/$\Delta\varepsilon^{178}$Hf=2.368 and $\Delta\varepsilon^{176}$Hf/$\Delta\varepsilon^{180}$Hf=-1.527. With ep/th in the range 0.44 to 2.2 documented by Sprung *et al.*[21], $\Delta\varepsilon^{176}$Hf/$\Delta\varepsilon^{178}$Hf ranges from 2.332 to 2.361 and $\Delta\varepsilon^{176}$Hf/$\Delta\varepsilon^{180}$Hf ranges from -1.542 to -1.530. The ratio $\Delta\varepsilon^{180}$Hf/$\Delta\varepsilon^{178}$Hf varies little with the fractions of thermal and epithermal neutrons (~-1.53 for ep/th ratios relevant to lunar samples), so this is not a basis to prefer either $\Delta\varepsilon^{178}$Hf or $\Delta\varepsilon^{180}$Hf for neutron-capture correction on $\varepsilon^{176}$Hf. The observed slope $\Delta\varepsilon^{180}$Hf/$\Delta\varepsilon^{178}$Hf in lunar samples is -1.58 (Sprung *et al.*[21], this study), which is close to our theoretically predicted slope of -1.53. For comparison, the calculations presented in Sprung *et al.*[21] yield $\Delta\varepsilon^{176}$Hf/$\Delta\varepsilon^{178}$Hf=2.613, $\Delta\varepsilon^{176}$Hf/$\Delta\varepsilon^{180}$Hf=-1.652, and $\Delta\varepsilon^{180}$Hf/$\Delta\varepsilon^{178}$Hf=-1.58. Estimating systematic errors introduced by model calculations of cosmogenic effects is difficult and we take the differences between the slopes inferred by Sprung *et al.*[21] and ours as a measure of uncertainty. We therefore have $\Delta\varepsilon^{176}$Hf/$\Delta\varepsilon^{178}$Hf = 2.35±0.25 (±11%) and $\Delta\varepsilon^{176}$Hf/$\Delta\varepsilon^{180}$Hf = -1.54±0.11 (±7%). The fact that model predictions match the measured $\Delta\varepsilon^{180}$Hf/$\Delta\varepsilon^{178}$Hf ratio within ~3% strengthens the view that model predictions are accurate. To further assess the reliability of the correction for cosmogenic effects on $^{176}$Hf, we apply both $\Delta\varepsilon^{178}$Hf and $\Delta\varepsilon^{180}$Hf corrections and compare the two (the *c* subscript stands for corrected for cosmogenic effects and *p* stands for present measured value),

$$\varepsilon^{176}\mathrm{Hf}_{\mathrm{zrc\text{-}p},c} = \varepsilon^{176}\mathrm{Hf}_{\mathrm{zrc\text{-}p}} - \alpha_i \varepsilon^i\mathrm{Hf}_{\mathrm{zrc}}. \quad (13)$$

We can also write this correction for absolute ratios using the approximation $\varepsilon^{176}\mathrm{Hf}_{\text{zrc-p},c} - \varepsilon^{176}\mathrm{Hf}_{\text{zrc-p}} \approx 10^4\left[\left({}^{176}\mathrm{Hf}/{}^{177}\mathrm{Hf}\right)^*_{\text{zrc-p},c}/\left({}^{176}\mathrm{Hf}/{}^{177}\mathrm{Hf}\right)^*_{\text{zrc-p}} - 1\right]$,

$$\left(\frac{^{176}\mathrm{Hf}}{^{177}\mathrm{Hf}}\right)^*_{\text{zrc-}p,c} = \left(\frac{^{176}\mathrm{Hf}}{^{177}\mathrm{Hf}}\right)^*_{\text{zrc-p}}\left(1 - \frac{\alpha_i \varepsilon^i\mathrm{Hf},}{10^4}\right), \tag{14}$$

where $\alpha_i = 2.35$ for $i = 178$ and $\alpha_i = -1.54$ for $i = 180$.

We also apply the correction given by Sprung *et al.*[21] and compare the results with our updated formulas (**Table 4**). The lunar zircons analyzed in this study have $\varepsilon^{178}\mathrm{Hf}$ values that vary between ~0 and 2, and $\varepsilon^{180}$Hf values that vary between ~0 and -2. **Table 4** shows how the different corrections compare with each other for our data set. The corrections on $\varepsilon^{176}\mathrm{Hf}$ using $\varepsilon^{178}\mathrm{Hf}$ or $\varepsilon^{180}\mathrm{Hf}$ and the model predictions from either Sprung *et al.*[21] or the present work show some scatter but we see no systematic offset between the corrections, suggesting that much of the uncertainty in the correction stems from the precision with which the cosmogenic shifts are measured.

Correction for decay of $^{176}$Lu into $^{176}$Hf after zircon crystallization is done by combining TIMS high-precision U-Pb crystallization ages and measured Lu/Hf ratios. TIMS U-Pb dating was done at Princeton University using established procedures [2, 77]. The U-Pb age was calculated for each leachate and trace elements (including Lu/Hf ratio) were analyzed on an aliquot of each solution using a Thermo iCAP single collector quadrupole ICP-MS following the methodology in O'Connor *et al.*[78]. Measurements were calibrated using a matrix-matched external calibration solution with Lu, Hf, and other trace elements proportional to that in natural zircon. Instrument drift and data reproducibility were monitored by measurement of four independent standard solutions: MUNZirc 1-2c, MUNZirc 3-2c, Plesovice[54, 79], and an in-house Zr-Hf standard. The initial zircon $^{176}$Hf/$^{177}$Hf ratio corrected for both cosmogenic effects and $^{176}$Lu decay, is,

$$\left(\frac{^{176}\mathrm{Hf}}{^{177}\mathrm{Hf}}\right)_{\mathrm{zrc-t,c}} = \left(\frac{^{176}\mathrm{Hf}}{^{177}\mathrm{Hf}}\right)_{\mathrm{zrc}\text{-}p}\left(1-\frac{\alpha_i\varepsilon^i\mathrm{Hf,}}{10^4}\right)-\left(\frac{^{176}\mathrm{Lu}}{^{177}\mathrm{Hf}}\right)_{\mathrm{zrc}\text{-}p}\left(e^{\lambda_{^{176}\mathrm{Lu}}t}-1\right), \quad (15)$$

The time *t* represents the crystallization age of the zircon given by isotopic closure of the U/Pb system, which is assumed equivalent to that for Lu/Hf. The main uncertainty with this correction lies in the measured Lu/Hf ratio as any systematic error in the decay constant does partially cancel out when the Hf isotopic composition is expressed relative to that of the chondritic uniform reservoir CHUR at the same time.

A potential concern with data interpretation is whether U-Pb and Lu-Hf systematics measured in a leachate or residue can be reliably linked or if they were decoupled either during residence on the Moon or chemical processing in the laboratory. The chemical abrasion procedure employed for U-Pb dating could have fractionated the Lu/Hf ratio by incongruent dissolution. Those two systems could also have been decoupled by zoning or Pb loss after metamictisation. One powerful approach for addressing this concern is to compare the calculated initial $\varepsilon^{176}$Hf values from successive leaching steps and residues. Indeed, zircons with simple single-stage histories are anticipated to exhibit uniform initial Hf isotopic composition. Variations in initial $\varepsilon^{176}$Hf values might indicate that the zircon had a complex multi-stage history, making it unsuitable for dating, or that there was incongruent Lu/Hf dissolution, rendering the age adjustment questionable. Several leaching steps were analyzed in three zircons: 14163 Z26 (L1, L2), 14321 Z3 (L1, L2), and 72275 Z1 (L1, L2, and R). The differences in raw $\varepsilon^{176}$Hf between the two leaching steps of 14163 Z26 and 14321 Z3 are 0.89±1.16 and 0.18±0.59 ε-units, respectively, which is indistinguishable from zero. The mean square weighted deviate on the three raw $^{176}$Hf/$^{177}$Hf ratios measured in 72275 Z1 R is 0.64, when the 2-sided 95% confidence interval for the reduced-$\chi^2$ distribution for $n = 3 - 1 = 2$ degrees of freedom is 0.025 to 3.69, indicating that the three values are identical within their given errors. If the cosmogenic correction is accurate, U-Pb ages have not been disturbed, and Lu and Hf dissolved congruently, we would expect initial $\varepsilon^{176}$Hf values to remain consistent within error. Not propagating the error on CHUR, the $\varepsilon^{176}$Hf values are +0.09±0.96 and -2.63±0.96 for 14163 Z26, -2.17±0.80 and -3.79±1.11 for 14321 Z3, and -0.57±1.02, -1.22±1.06, and -1.53±0.96 for 72275 Z1 R. A minor discrepancy exists between the two corrected values for 14163 Z26, which might stem from incongruent dissolution of Lu/Hf. All other values align with each other. Overall, our findings show no clear evidence of disturbance in coupled Lu-Hf and U-Pb systematics of analyzed zircons. However, given the limited number of measurements in our dataset, there remains possibilities that zircons with a complex history could exhibit such disturbances. Therefore, A thorough examination of the initial $\varepsilon^{176}$Hf values, Lu/Hf ratios, and U-Pb ages from different leachates is recommended to filter out zircons with complex histories.

The CHUR $^{176}$Hf/$^{177}$Hf evolution is most accurately calculated based on estimates of the $^{176}$Hf/$^{177}$Hf ratio at solar system formation (*ss*=Solar System Initial) of 0.279781±0.000018 and the present-day chondritic $^{176}$Lu/$^{177}$Hf ratio of 0.0338±0.0001 [25],

$$\left(\frac{^{176}\mathrm{Hf}}{^{177}\mathrm{Hf}}\right)_{\mathrm{CHUR-t}} = \left(\frac{^{176}\mathrm{Hf}}{^{177}\mathrm{Hf}}\right)_{\mathrm{CHUR}\text{-}ss} + \left(\frac{^{176}\mathrm{Lu}}{^{177}\mathrm{Hf}}\right)_{\mathrm{CHUR}\text{-}p} \left(e^{\lambda_{^{176}\mathrm{Lu}} t_{ss}} - e^{\lambda_{^{176}\mathrm{Lu}} t}\right), \quad (16)$$

where CHUR $-\ t$, CHUR-$ss$, and CHUR-$p$ denote CHUR composition at time $t$ before present, at the formation of the solar system, and at present, and $t_{ss}$ is the age of the solar system. We are interested in tracking how the initial $^{176}$Hf/$^{177}$Hf isotopic compositions of zircons compare relative to CHUR taken at the same time. Because isotopic variations are small, it is customary to express them in $\varepsilon^{176}$Hf notation $\varepsilon^{176}\mathrm{Hf}_t = \left[\left(^{176}\mathrm{Hf}/^{177}\mathrm{Hf}\right)_t/\left(^{176}\mathrm{Hf}/^{177}\mathrm{Hf}\right)_{\mathrm{STD}} - 1\right] \times 10^4$, where STD denotes a reference material. The standard is usually taken to be CHUR-$t$. Relative to this reference, the $\varepsilon^{176}\mathrm{Hf}_t$ value of zircons can be expressed as,

$$\varepsilon^{176}\text{Hf}_{\text{zrc-t,c/CHUR-t}} = \left[ \frac{\left(\frac{^{176}\text{Hf}}{^{177}\text{Hf}}\right)_{\text{zrc-p}} \left(1 - \frac{\alpha_i \varepsilon^i \text{Hf}}{10^4}\right) - \left(\frac{^{176}\text{Lu}}{^{177}\text{Hf}}\right)_{\text{zrc-}p} \left(e^{\lambda_{^{176}\text{Lu}} t} - 1\right)}{\left(\frac{^{176}\text{Hf}}{^{177}\text{Hf}}\right)_{\text{CHUR-}ss} + \left(\frac{^{176}\text{Lu}}{^{177}\text{Hf}}\right)_{\text{CHUR-}p} \left(e^{\lambda_{^{176}\text{Lu}} t_{ss}} - e^{\lambda_{^{176}\text{Lu}} t}\right)} - 1 \right] \times 10^4. \quad (17)$$

## Uncertainties

The uncertainties of $\varepsilon^{176}\text{Hf}_{\text{zrc-}t,c}$ were propagated using both analytical and Monte-Carlo methods. The analytical approach can be implemented in a spreadsheet, but it involves making some approximations that can be tested using the Monte-Carlo approach. The parameters in Eq. 17 that are uncertain are: the measured internally normalized $\left(^{176}\text{Hf}/^{177}\text{Hf}\right)_{\text{zrc-}p}$ ratio ($x_1$), the factor $\alpha_i$ used to correct cosmogenic effects $\alpha_{178} = 2.35 \pm 0.25$ and $\alpha_{180} = -1.54 \pm 0.11$ ($x_2$), the measured isotopic shifts $\varepsilon^i$Hf in $^{178}$Hf or $^{180}$Hf that are used to correct cosmogenic effects ($x_3$), the $\left(^{176}\text{Lu}/^{177}\text{Hf}\right)_{\text{zrc-}p}$ ratio of the zircon used to correct for in-situ decay of $^{176}$Lu ($x_4$), the decay constant $\lambda_{^{176}\text{Lu}} = 1.867 \pm 0.008 \times 10^{-11}$ Ga$^{-1}$ ($x_5$ [1]), the crystallization age of the zircon $t$ ($x_6$), the CHUR parameters $\left(^{176}\text{Hf}/^{177}\text{Hf}\right)_{\text{CHUR-}ss} = 0.279781 \pm 0.000018$ ( $x_7$ ) and $\left(^{176}\text{Lu}/^{177}\text{Hf}\right)_{\text{CHUR-}p} = 0.0338 \pm 0.0001$ ($x_8$), and the age of the solar system $t_{ss} = 4567.3 \pm 0.16$ Ma ($x_9$, [80]). As discussed earlier, some of these uncertainties like that on the decay constant largely cancel out when using the $\varepsilon^{176}$Hf notation normalized to contemporaneous CHUR, but we propagated all uncertainties to evaluate which ones could safely be neglected and we provide a simplified formula that only considers the ones that matter. Some assumptions/approximations must be made to derive an analytic equation, most notably that the functional relationship can be linearized over the range defined by uncertainties and that the distribution remains approximately gaussian. To evaluate the accuracy of the analytical approach, we have also run Monte Carlo simulations (MCS) by randomly generating a large number (200000) of multi-normal random values for $(x_1, x_2, x_3, x_4, x_5, x_6, x_7, x_8, x_9)$ based on the quoted values and uncertainties. All uncertainties are taken to be independent, except for the measured $\left({}^{176}\mathrm{Hf}/{}^{177}\mathrm{Hf}\right)_{\mathrm{zrc}\text{-}p}$ ratio ($x_1$) on the one hand, and $\varepsilon^{178}$Hf and $\varepsilon^{180}$Hf ($x_3$) used for correcting cosmogenic effects on the other hand. This correlation in errors arises from taking ratios of all isotopes to $^{177}$Hf and applying the same $^{179}$Hf/$^{177}$Hf internal normalization scheme to all ratios. The error correlations (correlation coefficient) are calculated based on cycle-level variations. The data not only involve internal normalization but also normalization to bracketing standards, which affects error correlations. This was accounted for using the formulas of Dauphas *et al.*[81]. Details are provided in the Supplementary materials. In the MCS, we use a joint binormal probability distribution to generate $(x_1, x_3)$. All other values are assumed to be independent. We also consider the correlation coefficient between $x_1$ and $x_3$ in the derivation of the analytical formula. The calculated correlation coefficients ($\rho$) are compiled in **Table S2**.

We used two forms of Eq. 17 to calculate the errors on CHUR-normalized $\varepsilon^{176}$Hf. The first one is Eq. 17 with the CHUR value a random variable in the denominator, $\varepsilon^{176}\mathrm{Hf}_{\mathrm{zrc\text{-}t,c/CHUR\text{-}t}} = f_1(x_1, x_2, x_3, x_4, x_5, x_6, x_7, x_8, x_9)$ with

$$f_1 = \left[\frac{x_1\left(1-\frac{x_2 x_3}{10^4}\right)-x_4\left(e^{x_5 x_6}-1\right)}{x_7+x_8\left(e^{x_5 x_9}-e^{x_5 x_6}\right)}-1\right]\times 10^4. \tag{18}$$

The second expression separates the errors arising from $^{176}$Hf/$^{177}$Hf measurements and CHUR parameters. The reasons for doing so are the following: (*i*) The analytical solution starting with the full expression (Eq. 18) would be cumbersome to derive and use. (*ii*) Equation 17 encompasses errors from both sample measurements and CHUR normalization, with the latter a systematic error that affects all zircons and creates statistical dependency between model ages. This is important if one is interested in examining the statistical distribution of zircon model ages.

(*iii*) Splitting CHUR and zircon uncertainties allows us to partly separate uncertainties from the literature and uncertainties tied to the quality of our measurements.

To split the errors from CHUR and the zircon measurement, we recognize that,

$$\varepsilon^{176}\mathrm{Hf}_{\text{zrc-t,c/CHUR-t}} = \left[\frac{\left({}^{176}\mathrm{Hf}/{}^{177}\mathrm{Hf}\right)_{\text{zrc-t,c}}}{\left({}^{176}\mathrm{Hf}/{}^{177}\mathrm{Hf}\right)_{\text{CHUR-t}}} - 1\right]10^4 \simeq \left[\frac{\left({}^{176}\mathrm{Hf}/{}^{177}\mathrm{Hf}\right)_{\text{zrc-t,c}}}{\left({}^{176}\mathrm{Hf}/{}^{177}\mathrm{Hf}\right)_{\text{STD}}} - 1\right]10^4 - \left[\frac{\left({}^{176}\mathrm{Hf}/{}^{177}\mathrm{Hf}\right)_{\text{CHUR-t}}}{\left({}^{176}\mathrm{Hf}/{}^{177}\mathrm{Hf}\right)_{\text{STD}}} - 1\right]10^4 \simeq \varepsilon^{176}\mathrm{Hf}_{\text{zrc-t/STD}} - \varepsilon^{176}\mathrm{Hf}_{\text{CHUR-t/STD}}, \tag{19}$$

where STD can be any standard of our choosing. While Eqs. 18 and 19 yield the same values for $\varepsilon^{176}\mathrm{Hf}_{\text{zrc-t,c/CHUR-}t}$, the two formulas lead to handling error propagation differently. When propagating errors, Eq. 17 would treat $X = \left({}^{176}\mathrm{Hf}/{}^{177}\mathrm{Hf}\right)_{zrc\text{-}t}$ and $Y = \left({}^{176}\mathrm{Hf}/{}^{177}\mathrm{Hf}\right)_{CHUR\text{-}t}$ as random variables in $\varepsilon = (X/Y - 1)10^4$ and would introduce dependency in calculated model ages. If we use Eq. 17 and adopt for the standard the exact value of CHUR $\tilde{Y}$ (the tilde accent is here to indicate that it is not a random variable) at the time of the zircon crystallization, $\varepsilon \simeq \left(X/\tilde{Y} - 1\right)10^4 - \left(Y/\tilde{Y} - 1\right)10^4$, we separate the components of the uncertainty that lead to inter-dependency in model ages of individual zircons. In this approach, we have,

$$\varepsilon^{176}\mathrm{Hf}_{\text{zrc-t,c/}\widetilde{\text{CHUR}}\text{-t}} = \left[\frac{\left(\frac{{}^{176}\mathrm{Hf}}{{}^{177}\mathrm{Hf}}\right)_{\text{zrc-p}}\left(1-\frac{\alpha_i\varepsilon^i\mathrm{Hf}}{10^4}\right)-\left(\frac{{}^{176}\mathrm{Lu}}{{}^{177}\mathrm{Hf}}\right)_{\text{zrc-}p}\left(e^{\lambda_{^{176}\mathrm{Lu}}t}-1\right)}{\left(\widetilde{\frac{{}^{176}\mathrm{Hf}}{{}^{177}\mathrm{Hf}}}\right)_{\text{CHUR-}ss}+\left(\widetilde{\frac{{}^{176}\mathrm{Lu}}{{}^{177}\mathrm{Hf}}}\right)_{\text{CHUR-}p}\left(e^{\tilde{\lambda}_{^{176}\mathrm{Lu}}\tilde{t}_{ss}}-e^{\tilde{\lambda}_{^{176}\mathrm{Lu}}\tilde{t}}\right)} - 1\right] \times 10^4. \tag{20}$$

$$\varepsilon^{176}\mathrm{Hf}_{\text{CHUR-t/}\widetilde{\text{CHUR}}\text{-t}} = \left[\frac{\left(\frac{{}^{176}\mathrm{Hf}}{{}^{177}\mathrm{Hf}}\right)_{\text{CHUR-}ss}+\left(\frac{{}^{176}\mathrm{Lu}}{{}^{177}\mathrm{Hf}}\right)_{\text{CHUR-}p}\left(e^{\lambda_{^{176}\mathrm{Lu}}t_{ss}}-e^{\lambda_{^{176}\mathrm{Lu}}t}\right)}{\left(\widetilde{\frac{{}^{176}\mathrm{Hf}}{{}^{177}\mathrm{Hf}}}\right)_{\text{CHUR-}ss}+\left(\widetilde{\frac{{}^{176}\mathrm{Lu}}{{}^{177}\mathrm{Hf}}}\right)_{\text{CHUR-}p}\left(e^{\tilde{\lambda}_{^{176}\mathrm{Lu}}\tilde{t}_{ss}}-e^{\tilde{\lambda}_{^{176}\mathrm{Lu}}\tilde{t}}\right)} - 1\right] \times 10^4\ , \tag{21}$$

where variables with tilde are exact values with no error. The expected value of $\varepsilon^{176}\mathrm{Hf}_{\text{CHUR-t/}\widetilde{\text{CHUR}}\text{-t}}$ is 0 at all ages. The denominator is no longer a random variable and we note it as $C = \left(\widetilde{{}^{176}\mathrm{Hf}/{}^{177}}\mathrm{Hf}\right)_{\text{CHUR-}ss} + \left(\widetilde{{}^{176}\mathrm{Hf}/{}^{177}}\mathrm{Hf}\right)_{\text{CHUR-}p}\left(e^{\tilde{\lambda}_{^{176}\mathrm{Lu}}\tilde{t}_{ss}} - e^{\tilde{\lambda}_{^{176}\mathrm{Lu}}\tilde{t}}\right)$. Equations 20-21 can be written as functions $f_2(x_1, x_2, x_3, x_4, x_5, x_6)$ and $f_3(x_5, x_6, x_7, x_8, x_9)$ using the variables mentioned above,

$$\varepsilon^{176}\text{Hf}_{\text{zrc-t,c/}\widetilde{\text{CHUR}}\text{-t}} = f_2 = \left[\frac{x_1 - x_4(e^{x_5 x_6} - 1)}{C} - 1\right] \times 10^4 - \frac{x_1 x_2 x_3}{C}, \tag{22}$$

$$\varepsilon^{176}\text{Hf}_{\text{CHUR-t/}\widetilde{\text{CHUR}}\text{-t}} = f_3 = \left[\frac{x_7 + x_8(e^{x_5 x_9} - e^{x_5 x_6})}{C} - 1\right] \times 10^4. \tag{23}$$

We note,

$$\varepsilon^{176}\text{Hf}_{\text{zrc-t,c/}\widetilde{\text{CHUR}}\text{-t}} - \varepsilon^{176}\text{Hf}_{\text{CHUR-t/}\widetilde{\text{CHUR}}\text{-t}} = f_4 = \frac{x_1 - x_4(e^{x_5 x_6} - 1) - x_7 - x_8(e^{x_5 x_9} - e^{x_5 x_6})}{C} \times 10^4 - \frac{x_1 x_2 x_3}{C}. \tag{24}$$

The uncertainties of the various parameters in Eqs. 22, 23 and 24 can be propagated analytically using the approximation, $\sigma_f^2 \simeq \sum \left(\frac{\partial f}{\partial x_i}\right)^2 \sigma_{x_i}^2 + 2\sum\sum \frac{(\partial f)^2}{\partial x_i \partial x_j} \sigma_{x_i x_j}$, where $\sigma_{x_i x_j}$ is the covariance (see Supplementary Materials for details). A virtue of the analytical approach is that it is easy to quantify the contribution of each variable to the total variance (**Table S1**). Examination of the error budget (**Table S1**) shows that some uncertainties can be safely neglected, and we can set $\left(\frac{\partial f}{\partial x_i}\right)^2 \sigma_{x_i}^2$ for variables $x_2$,$x_4$,$x_5$,$x_6$,$x_8$, and $x_9$ to 0 without losing too much accuracy in error estimates. Under those conditions, we have,

$$\sigma_{f_4}^2 \simeq \left(\frac{10^4}{C}\right)^2 \left[\left(1 - \frac{x_2 x_3}{10^4}\right)^2 \sigma_{x_1}^2 + \sigma_{x_7}^2\right] + \frac{x_1^2 x_2^2}{C^2} \sigma_{x_3}^2 + 2\frac{10^4}{C^2}\left(1 - \frac{x_2 x_3}{10^4}\right) x_1 x_3 \sigma_{x_1 x_3}. \tag{25}$$

These formulas are implemented in an Excel spreadsheet provided as Table S2 that can be used for data reduction.

**Figure 8** shows the results of the five lunar zircons measured here. The final uncertainty of each $\varepsilon^{176}\text{Hf}_{\text{zrc-}t,c}$ value is calculated by propagating errors resulting from measurements of $^{176}$Hf/$^{177}$Hf, $^{178}$Hf/$^{177}$Hf, $^{180}$Hf/$^{177}$Hf, $^{176}$Lu/$^{177}$Hf, and the crystallization age *t*. In panels A andC, the uncertainties on $\varepsilon^{176}\text{Hf}_{\text{zrc-t,c/CHUR-t}}$ were calculated using Eq. 18. The uncertainty ellipse is slanted because of the different age-dependences of $\varepsilon^{176}\mathrm{Hf}_{\mathrm{zrc}\text{-}t,c}$ and $\varepsilon^{176}\mathrm{Hf}_{\mathrm{CHUR}}$. In this approach, distributions of zircon model ages are potentially less straightforward to interpret as the values of $x_5 = \lambda$, $x_7 = \left({}^{176}\mathrm{Hf}/{}^{177}\mathrm{Hf}\right)_{\mathrm{CHUR\text{-}ss}}$, $x_8 = \left({}^{176}\mathrm{Lu}/{}^{177}\mathrm{Hf}\right)_{\mathrm{CHUR}\text{-}p}$, and $x_9 = t_{\mathrm{ss}}$ will affect all zircon $\varepsilon^{176}\mathrm{Hf}_{\mathrm{zrc}\text{-}t,c} - \varepsilon^{176}\mathrm{Hf}_{\mathrm{CHUR}}$ analyses and the data points are not truly independent. In panels B and D, the uncertainties on $\varepsilon^{176}\mathrm{Hf}_{\mathrm{zrc}\text{-}t,c}$ and $\varepsilon^{176}\mathrm{Hf}_{\mathrm{CHUR}}$ are kept separated through the use of Eqs. 22 and 23. In this approach, the dependence of all individual zircon model ages on shared parameters $x_5$, $x_7$, $x_8$, and $x_9$ can be considered to calculate model age distributions of zircon populations. The difficulty with this approach is that the dependence of $\varepsilon^{176}\mathrm{Hf}_{\mathrm{zrc}\text{-}t,c}$ and $\varepsilon^{176}\mathrm{Hf}_{\mathrm{CHUR}}$ on crystallization age is lost and the individual uncertainty ellipses are less realistic. Both approaches to propagate uncertainties have some pitfalls that can be eliminated in a MCS by randomizing the shared parameters $x_5$, $x_7$, $x_8$, and $x_9$ for the entire data set, but as discussed below the two approaches yield almost identical age distributions. Our preferred option is to split uncertainties on $\varepsilon^{176}\mathrm{Hf}_{\mathrm{zrc}\text{-}t,c}$ and $\varepsilon^{176}\mathrm{Hf}_{\mathrm{CHUR}}$ because (*i*) it better portrays uncertainties arising from our zircon analyses rather than combining those with uncertainties on CHUR and the decay constant from the literature and (*ii*) an analytical expression is available to propagate uncertainties.

The main purpose of estimating the initial Hf isotopic compositions of zircons and comparing them with contemporaneous CHUR is to derive model ages of magmatic differentiation. Those model ages are calculated assuming a 2-stage evolution, where a reservoir R evolves with CHUR composition from $t_{ss}$ to $t_d$ before present, and then evolves with a fractionated $\left({}^{176}\mathrm{Lu}/{}^{177}\mathrm{Hf}\right)_{\mathrm{R}\text{-}p}$ ratio ($x_{10}$, the $p$ subscript indicates that the ${}^{176}\mathrm{Lu}/{}^{177}\mathrm{Hf}$ ratio of that hypothetical reservoir is expressed as what would be its present-day value) until the zircon crystallizes at $t$. In that context, the Hf isotopic composition of zircon and reservoir $R$ at time $t$ before present should be the same,

$$\left(\frac{^{176}\text{Hf}}{^{177}\text{Hf}}\right)_{\text{zrc-t,c}} = \left(\frac{^{176}\text{Hf}}{^{177}\text{Hf}}\right)_{R\text{-t}} = \left(\frac{^{176}\text{Hf}}{^{177}\text{Hf}}\right)_{\text{R-td}} + \left(\frac{^{176}\text{Lu}}{^{177}\text{Hf}}\right)_{R\text{-}p}\left(e^{\lambda_{^{176}\text{Lu}}t_d} - e^{\lambda_{^{176}\text{Lu}}t}\right). \quad (26)$$

The Hf isotopic composition for CHUR at time $t$ is,

$$\left(\frac{^{176}\text{Hf}}{^{177}\text{Hf}}\right)_{\text{CHUR-t}} = \left(\frac{^{176}\text{Hf}}{^{177}\text{Hf}}\right)_{\text{CHUR-}t_d} + \left(\frac{^{176}\text{Lu}}{^{177}\text{Hf}}\right)_{\text{CHUR-}p}\left(e^{\lambda_{^{176}\text{Lu}}t_d} - e^{\lambda_{^{176}\text{Lu}}t}\right). \quad (27)$$

Using the relationship $\left(^{176}\text{Hf}/^{177}\text{Hf}\right)_{\text{R-td}} = \left(^{176}\text{Hf}/^{177}\text{Hf}\right)_{\text{CHUR-td}}$, Eqs. 26 and 27 can be combined into the following equation,

$$\left(\frac{^{176}\text{Hf}}{^{177}\text{Hf}}\right)_{\text{zrc-t,c}} = \left(\frac{^{176}\text{Hf}}{^{177}\text{Hf}}\right)_{\text{CHUR-t}} + \left[\left(\frac{^{176}\text{Lu}}{^{177}\text{Hf}}\right)_{\text{R-}p} - \left(\frac{^{176}\text{Lu}}{^{177}\text{Hf}}\right)_{\text{CHUR-}p}\right]\left(e^{\lambda_{^{176}\text{Lu}}t_d} - e^{\lambda_{^{176}\text{Lu}}t}\right). \quad (28)$$

After some rearrangements, the model age $t_d$ can be expressed as,

$$t_d = \frac{1}{\lambda_{^{176}\text{Lu}}}\ln\left[e^{\lambda_{^{176}\text{Lu}}t} + \frac{\left(^{176}\text{Hf}/^{177}\text{Hf}\right)_{\text{CHUR-t}} - \left(^{176}\text{Hf}/^{177}\text{Hf}\right)_{\text{zrc-}t,c}}{\left(^{176}\text{Lu}/^{177}\text{Hf}\right)_{\text{CHUR-}p} - \left(^{176}\text{Lu}/^{177}\text{Hf}\right)_{\text{R-}p}}\right]. \quad (29)$$

For KREEP, $^{176}$Lu/$^{177}$Hf ratios of 0.0164 [10], 0.0154±0.0034[21] and 0.0153±0.0033[22] were used by previous authors. Higher $\left(^{176}\text{Lu}/^{177}\text{Hf}\right)_{\text{R-}p}$ ratios result in higher model ages. A minimum model age can be calculated taking $\left(^{176}\text{Lu}/^{177}\text{Hf}\right)_{\text{R-}p} = 0$ in Eq. 29,

$$t_d > \frac{1}{\lambda_{^{176}\text{Lu}}}\ln\left[e^{\lambda_{^{176}\text{Lu}}t} + \frac{\left(^{176}\text{Hf}/^{177}\text{Hf}\right)_{\text{CHUR-t}} - \left(^{176}\text{Hf}/^{177}\text{Hf}\right)_{\text{zrc-}t,c}}{\left(^{176}\text{Lu}/^{177}\text{Hf}\right)_{\text{CHUR-}p}}\right]. \quad (30)$$

The uncertainties for individual best estimate or minimum model ages are also evaluated using both MCS and analytical approaches. The analytical equations for propagating the uncertainties of model ages are provided in supplementary materials, and the two methods give the same errors for the model ages (**Fig. S1**). An initial $^{176}$Lu/$^{177}$Hf ratio of 0.0153±0.0033 was used to estimate model ages for lunar zircons and the results are compiled in **Table 5**.

Another way to calculate a model age is to do a linear regression of $\varepsilon^{176}\text{Hf}_{\text{zrc-}t,c}$ versus the crystallization age $(t)$ of all or a subset of zircons. This approach assumes that the population of zircons considered were all crystallized from an enriched reservoir that was isolated from CHUR at a set time. The intersection between the zircon regression line and CHUR gives the model age and the slope reflects the $\left({}^{176}\text{Lu}/{}^{177}\text{Hf}\right)_{\text{R-}p}$ of the reservoir R. As an illustration, we used a Monte Carlo approach to calculate the intercept model age and $\left({}^{176}\text{Lu}/{}^{177}\text{Hf}\right)_{\text{R-}p}$ ratio for the set of lunar zircons analyzed here. We generated a large random dataset (200000) for both zircon ages and $\varepsilon^{176}$Hf following previous procedure for each zircon, and then sampled each zircon data point from the synthetic dataset for a least-square fitting (grey lines in **Fig. 9 and 10**). The intersection with CHUR (orange points in **Fig. 9 and 10**) was calculated by either using Eqs. 22 and 23 and treating CHUR as constant ($\varepsilon^{176}\text{Hf}_{\text{CHUR-t}/\widehat{\text{CHUR}}\text{-t}} = 0$) or using Eq. 18 and treating CHUR as a variable simulated using a gaussian random number generator. In both cases, we calculated the slope and intersection with the CHUR line. The results of the two approaches are almost identical, yielding a model age of $4388^{-48}_{+76}$ Ma (95% C.I) and a $\left({}^{176}\text{Lu}/{}^{177}\text{Hf}\right)_{\text{R-}p}$ ratio of 0.0079 ± 0.0098 when $\varepsilon^{178}$Hf is used for correcting neutron capture effects. The model age and initial $\left({}^{176}\text{Lu}/{}^{177}\text{Hf}\right)_{\text{R-}p}$ ratio are $4344^{-38}_{+48}$ Ma (95% C.I) and 0.0082 ± 0.0098 when $\varepsilon^{180}$Hf is used for correcting neutron capture effects. The reason why we analyzed a small set of lunar zircons was to test our measurement protocol and data reduction pipeline on real samples. The zircon population analyzed here is too small to draw any conclusion about the history of LMO differentiation.

## Conclusion

A procedure was developed for the purification and separation of Zr and Hf from zircons. The focus of this procedure was on the isotopic analysis of Hf in single zircon grains but separation of Zr from Hf opens the door to isotopic analysis of Zr on the same samples. Hafnium isotopic analyses were done on MC-ICPMS, and we show that the precisions achieved are close to the theoretically attainable limit set by counting statistics and Johnson noise. We applied this technique to small zircon grains extracted from lunar samples returned by the Apollo missions. These zircon grains were chemically abraded and dated by U-Pb geochronology. The leachates from the chemical abrasion were passed though U-Pb chemistry before Hf purification. The zircon Hf isotopic analyses are corrected for *in situ* decay of $^{176}$Lu, neutron cature effects associated with exposure to cosmic rays in space, and are reported relative to chondrites. Analysis of these zircons allows us to test all aspects of our measurement protocol and data reduction pipeline, which we will apply to more lunar samples to refine our understanding of lunar impact and differentiation history.

**Acknowledgments**. We thank two anonymous reviewers for their constructive comments that greatly improved the quality of the manuscript. This work was supported by NASA grant 80NSSC20K0821 (EW) to KMcK, ND, MB, and BS, as well as grants NNX17AE86G (LARS), 80NSSC17K0744 (HW), 000306-002 (HW), 80NSSC21K0380 (EW), NSF grant EAR-2001098 (CSEDI), and funding from DOE to N.D. and by the Swiss National Science Foundation (IL, 200020_196955).

# Supporting Information

- Calculation of zircon initial $\varepsilon^{176}$Hf values using measured U-Pb ages, Lu/Hf ratios, neutron capture effects on $\varepsilon^{178}$Hf and $\varepsilon^{180}$Hf, and error propagation.
- Calculation of model ages of differentiation and error propagation.
- Excel spreadsheet for reduction of zircon Lu-Hf data.

# Figures and Tables

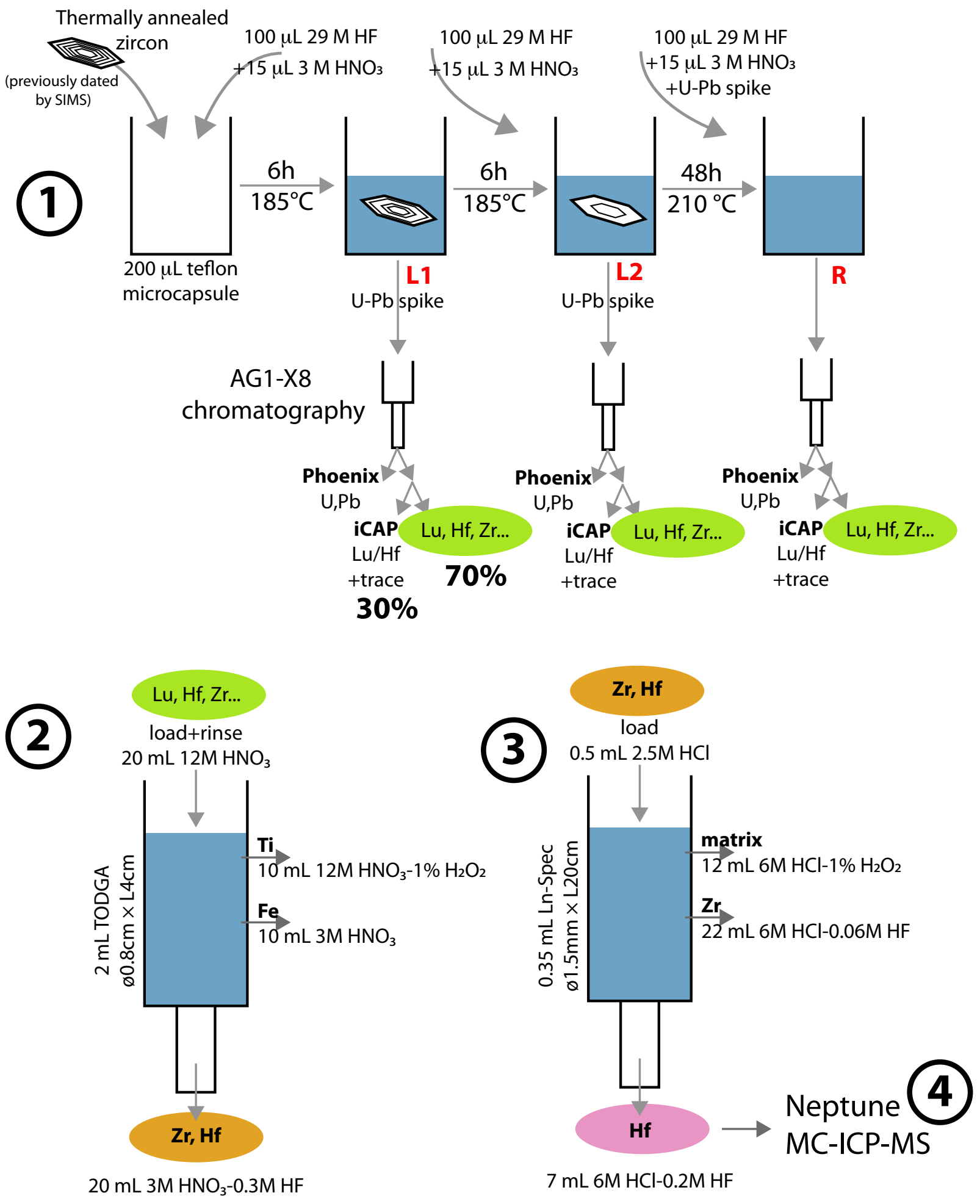

**Fig. 1.** Flowchart of the procedure used for Lu-Hf analysis of extraterrestrial zircons. 1. The zircons are thermally annealed and then subjected to chemical abrasion[2, 69]. Two leachates (L1 and L2) and one residue (R) solutions are retrieved for further processing. The solutions are passed on chromatography columns filled with AG1-X8 resin to separate U+Pb from a solution containing Lu, Hf, Zr, and other trace elements. The U+Pb elutions are analyzed using a Phoenix TIMS at Princeton University. The solution containing Lu, Hf, Zr, and other trace elements is split into two, with 30% used for Lu/Hf determination using an iCAP at Princeton University, and 70% used for Hf purification and Hf isotopic analysis at the Origins Lab of the University of Chicago. 2. The "Lu, Hf, Zr" 70% split solution is passed on a column filled with TODGA resin [52, 53] to separate Zr+Hf from most other elements. 3. The Zr+Hf cut is passed on a second column filled with Ln-Spec resin (Fig. 2) to separate Hf from Zr [27, 50]. 4. The isotopic composition of purified Hf is analyzed with a Neptune MC-ICP-MS at the University of Chicago.

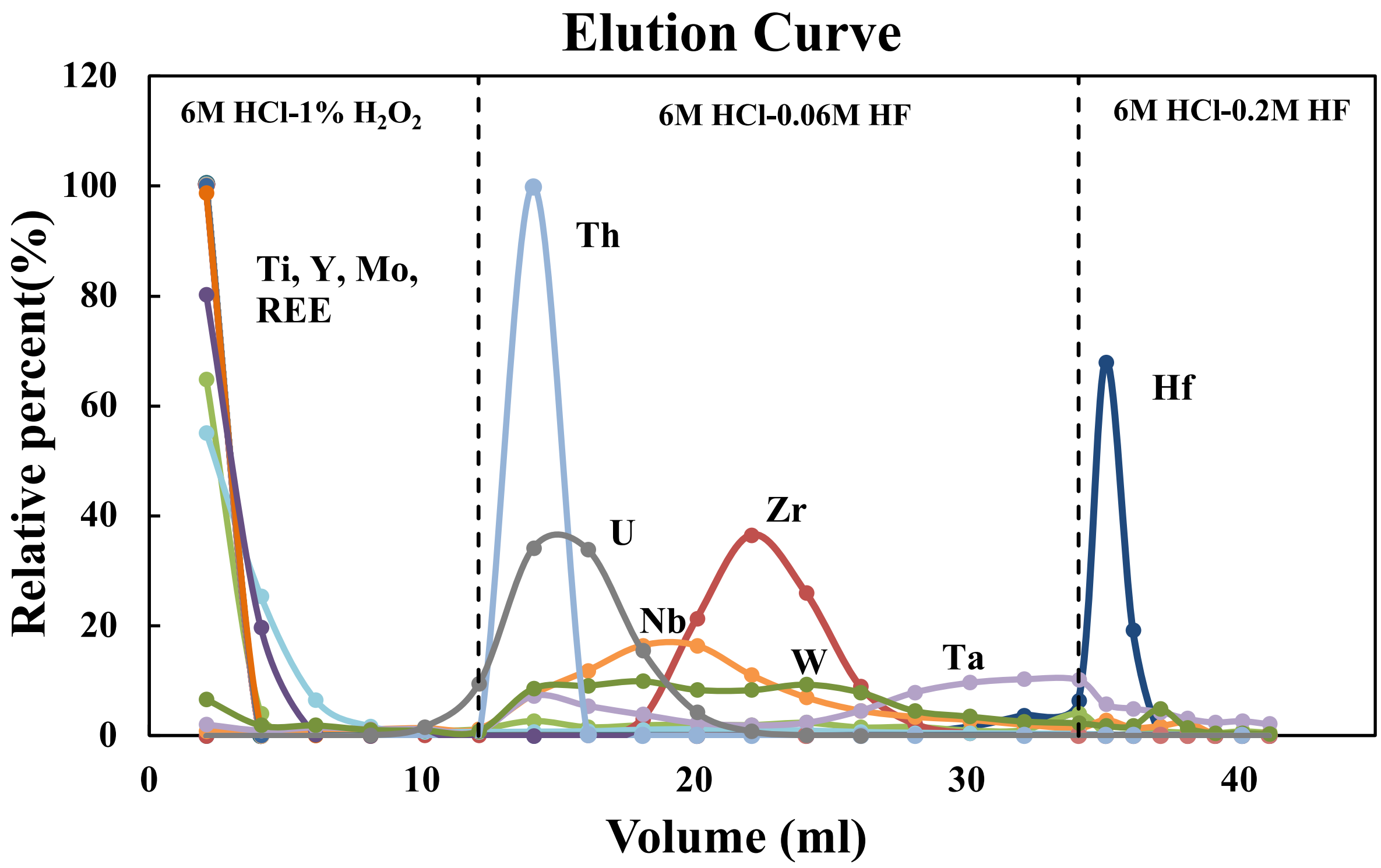

**Figure 2.** Elution curve of a multi-element standard solution on a 0.35 mL Teflon column loaded with Ln-Spec resin. This corresponds approximately to the second step of our Hf purification procedure. Matrix elements were removed with 12 mL of 6 M HCL – 1 vol% $H_2O_2$. Zirconium was first eluted in 22 mL of 6 M HCL - 0.06M HF. Hafnium was finally eluted in 7 mL of 6 M HCL - 0.2 M HF. The elution sequence is from Zhang *et al.*[50].

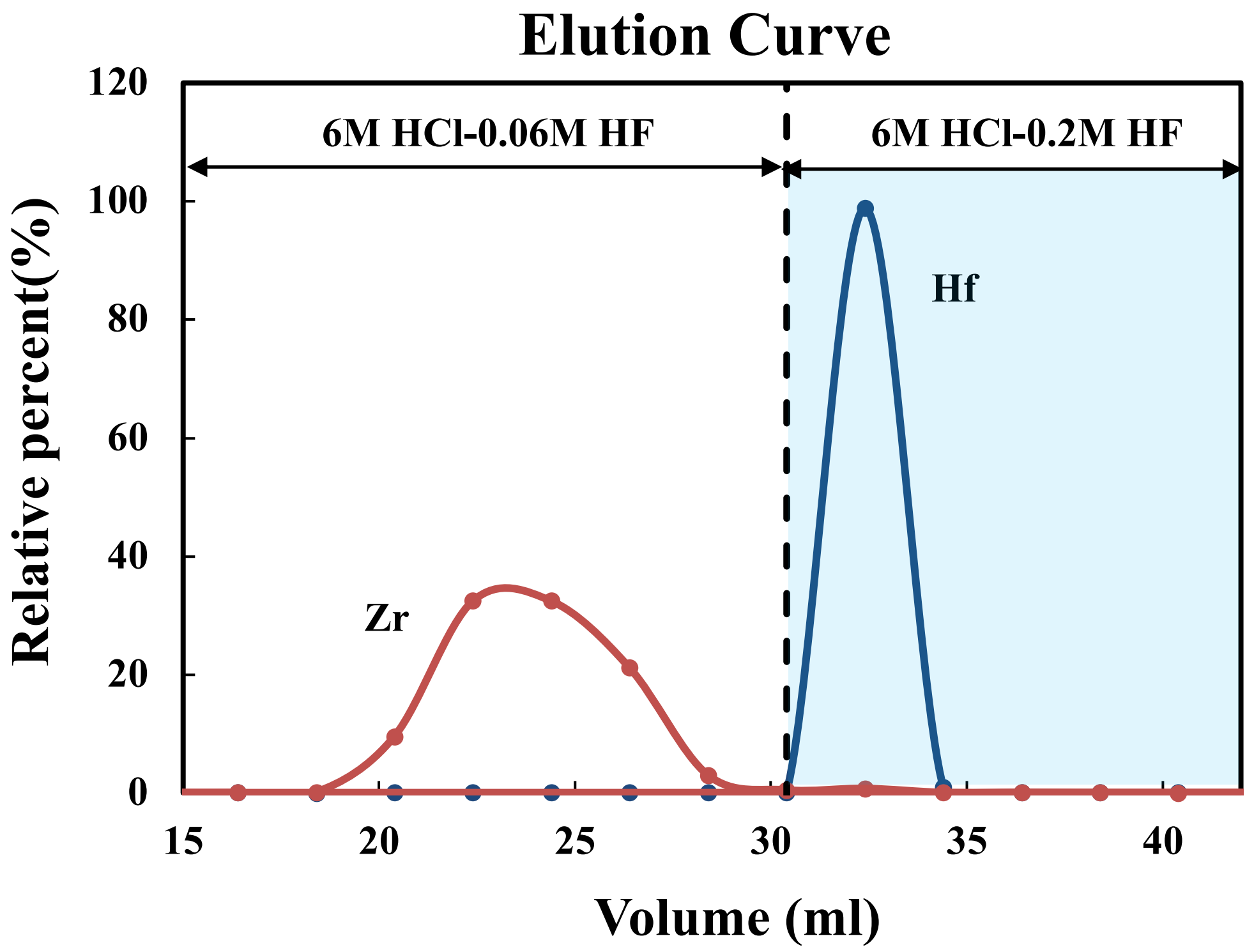

**Figure 3.** Elution curve of terrestrial zircon standard AS3 retrieved after U-Pb chemistry on a 0.35 mL Teflon column loaded with Ln-Spec resin. Matrix elements were removed with 12 mL of 6 M HCL – 1vol % $H_2O_2$. Zirconium was first eluted in 22 mL of 6 M HCL - 0.06 M HF. Hafnium was finally eluted in 7 mL of 6 M HCl - 0.2 M HF.

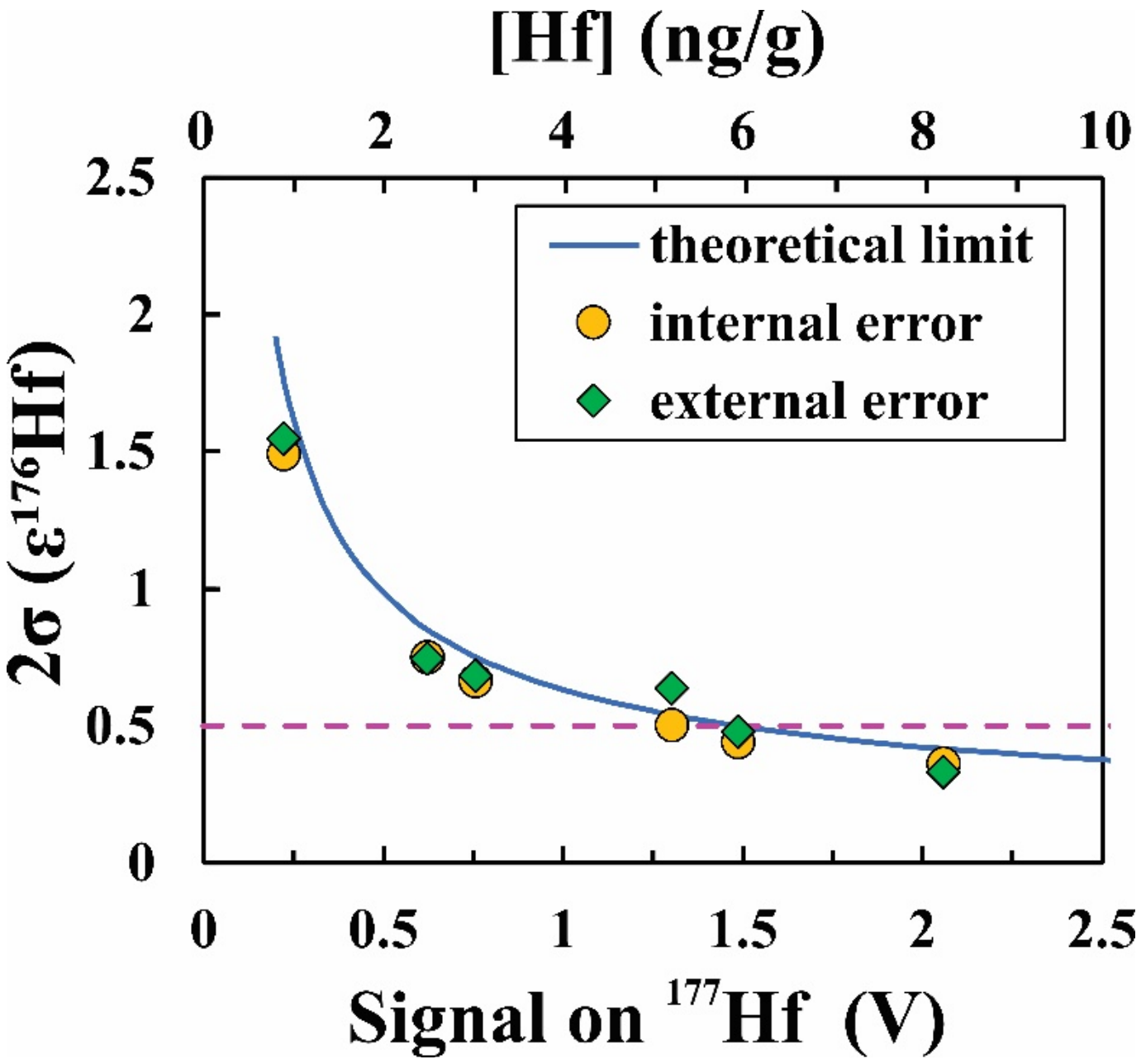

**Figure 4.** Measurement uncertainties (2σ) of $\varepsilon^{176}$Hf as a function of $^{177}$Hf signal intensity (bottom x-axis) and corresponding Hf concentration in the measured solutions (top x-axis) at a sensitivity of 0.22V for $^{177}$Hf (18.60% isotopic abundance) per ppb Hf with a sample uptake rate of ~100 μL/min measured for 10 min. The solid blue line is the theoretically achievable precision on $\varepsilon^{176}$Hf considering counting statistics and Johnson noise (Eq. 8; Dauphas *et al.*[60]). The internal and external errors from actual measurements agree well with the theoretically achievable precisions, which demonstrates that the current instrumental setup is optimized for Hf isotope measurements. To achieve a precision of better than ~±0.5 on $\varepsilon^{176}$Hf (dashed line) requires analysis of 20 ng Hf, which corresponds to a zircon of 94 μm in diameter.

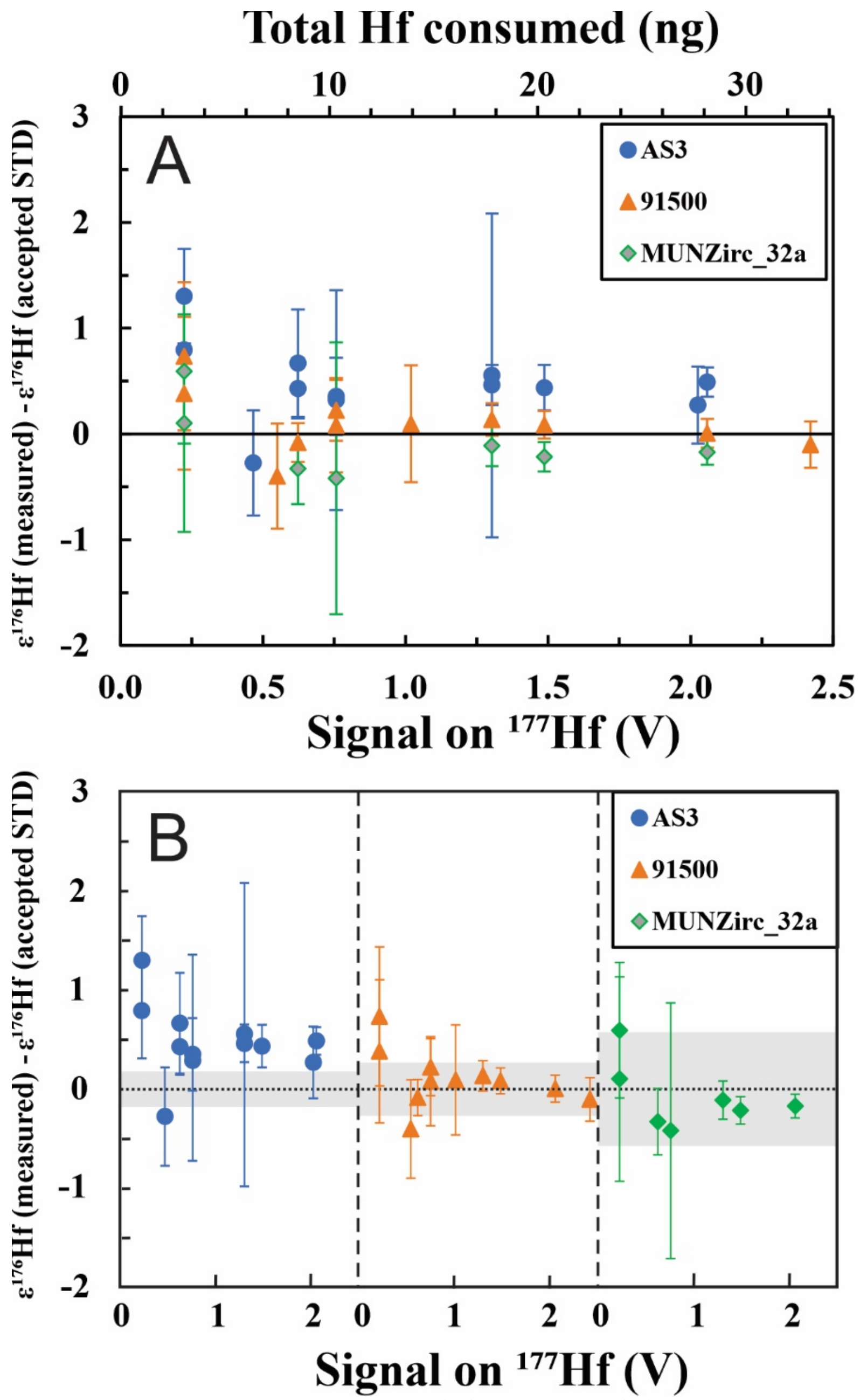

**Figure 5.** Relative difference from the literature $\varepsilon^{176}$Hf values[54, 61] of three zircon references (AS3, 91500, and MUNZirc 32a) as a function of $^{177}$Hf signal intensity (Table 2). The error bars in panel A are using our measurement errors only. Panel B shows those errors together with those reported for the reference values in the literature (grey bands). There is good agreement between measured and recommended $\varepsilon^{176}$Hf values at all solution Hf concentrations (signal on $^{177}$Hf), except possibly for AS3 but the literature value is for FC1, which are zircons from the same geological unit but extracted and processed at a different time. Each data point corresponds to a whole-chemistry replicate (involving the whole purification procedure, starting from the same solution but processing different amounts of Hf through chromatography).

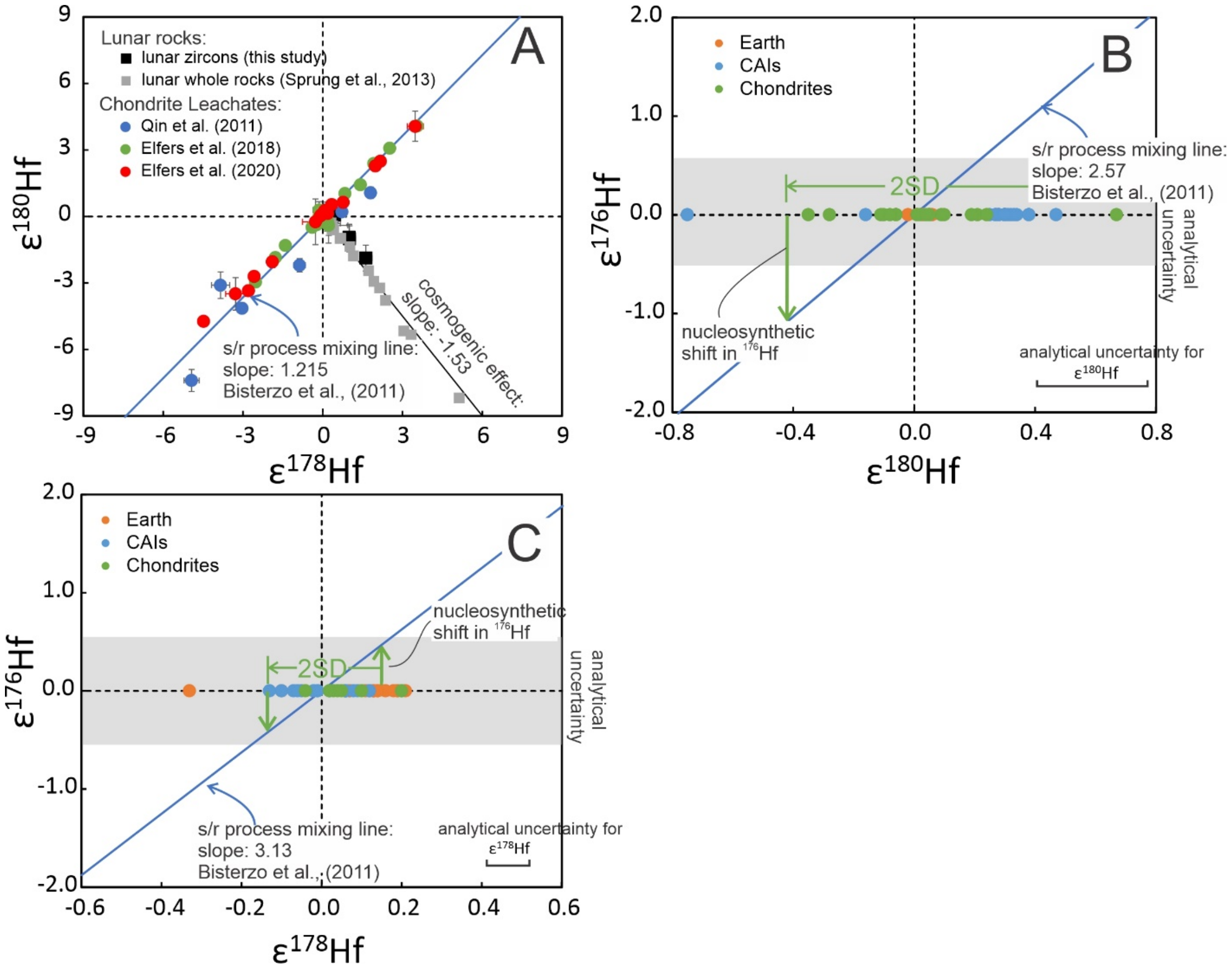

**Figure 6.** Nucleosynthetic isotopic variation of Hf in chondrites, chondrite leachates, CAIs, and terrestrial rocks. Panel A shows the $\varepsilon^{178}$Hf vs. $\varepsilon^{180}$Hf of chondrite leachates[42, 64, 65] and lunar zircons (this study) and lunar whole rocks[21]. The chondrite leachates follow the s/r-process mixing line with a positive slope[68], while lunar rocks sit on an almost perpendicular trend of cosmogenic effect with a negative slope. To constrain the nucleosynthesis effect on $\varepsilon^{176}$Hf, the $\varepsilon^{180}$Hf and $\varepsilon^{178}$Hf values of chondrites, CAIs and terrestrial rocks are projected onto the s/r process mixing line. Bulk chondrites show no resolvable nucleosynthetic anomalies in $\varepsilon^{180}$Hf (B) and $\varepsilon^{178}$Hf (C), which limits heterogeneities of nucleosynthetic origin on $\varepsilon^{176}$Hf in bulk planetary objects to less than ~±0.3.

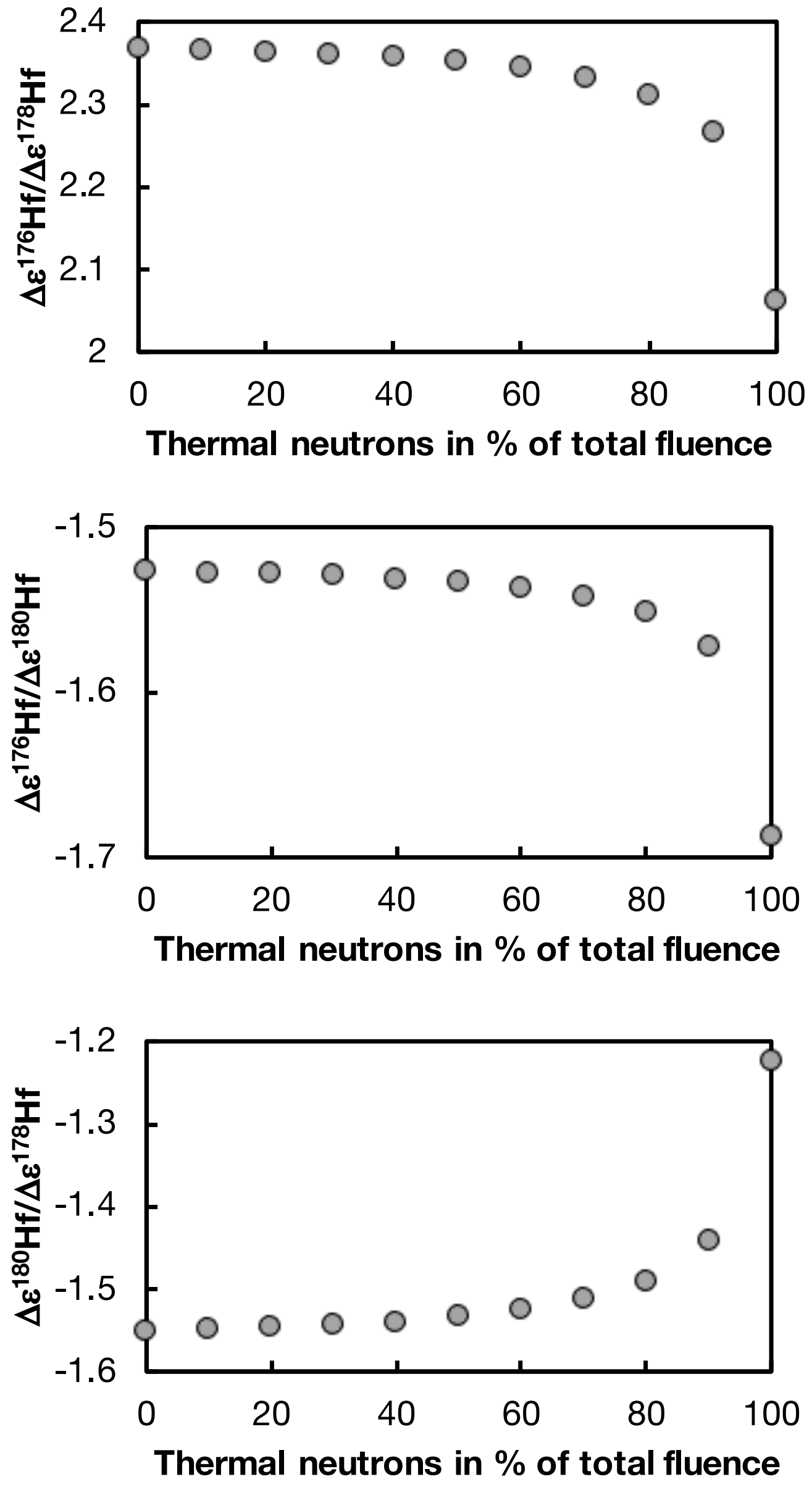

**Figure 7.** Ratio of neutron capture induced Hf isotopic shifts $\Delta\varepsilon^{176}$Hf/ $\Delta\varepsilon^{178}$Hf, $\Delta\varepsilon^{176}$Hf/ $\Delta\varepsilon^{180}$Hf and $\Delta\varepsilon^{180}$Hf/$\Delta\varepsilon^{178}$Hf as a function of fraction of thermal neutrons in the total neutron fluence. Previous work has shown that the proportion of thermal neutrons is 30 -70 % of the total fluence[21].

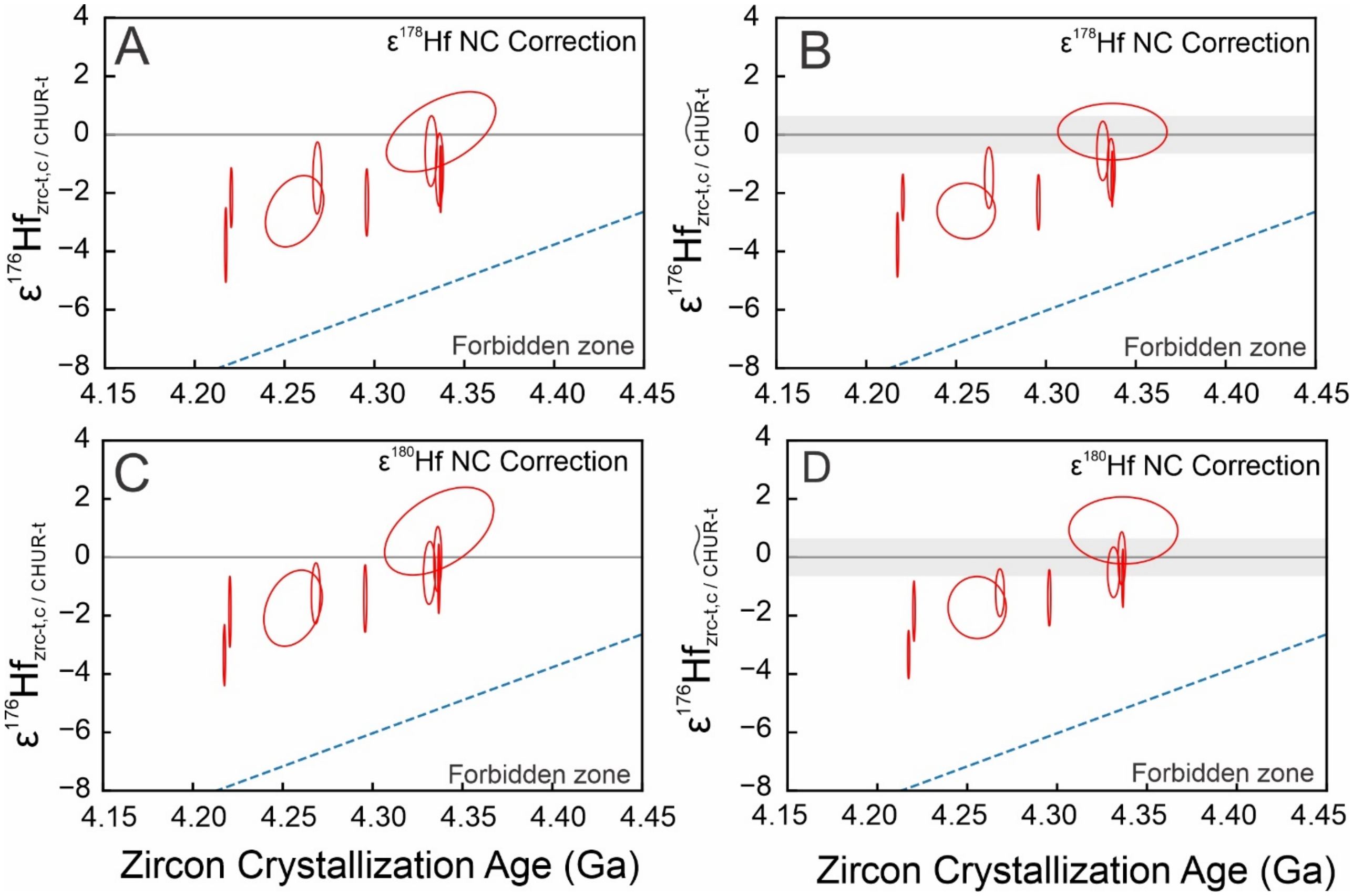

**Figure 8.** Single zircon initial $^{176}Hf/^{177}Hf$ isotopic composition (expressed as departure from CHUR in ε-unit) as a function of $^{207}Pb/^{206}Pb$ zircon crystallization age calculated using Eq. 18 (A, C) and Eqs. 22 and 23 (B, D). The difference between these two approaches is whether CHUR uncertainties are kept separated (grey bands in panels B, D) or not (panels A, C); see text for details. Each zircon $\varepsilon^{176}Hf_{Zrc-t,c/CHUR-t}$ value in this study was corrected for neutron capture effects using either $\varepsilon^{178}$Hf (A, B) or $\varepsilon^{180}$Hf (C, D) and the theoretically predicted cosmogenic effect correlation. Error ellipses are of 95% confidence level. The dashed line corresponds to a two-stage model evolution for a reservoir with Lu/Hf=0 isolated from the solar system at 4.567 Ga.

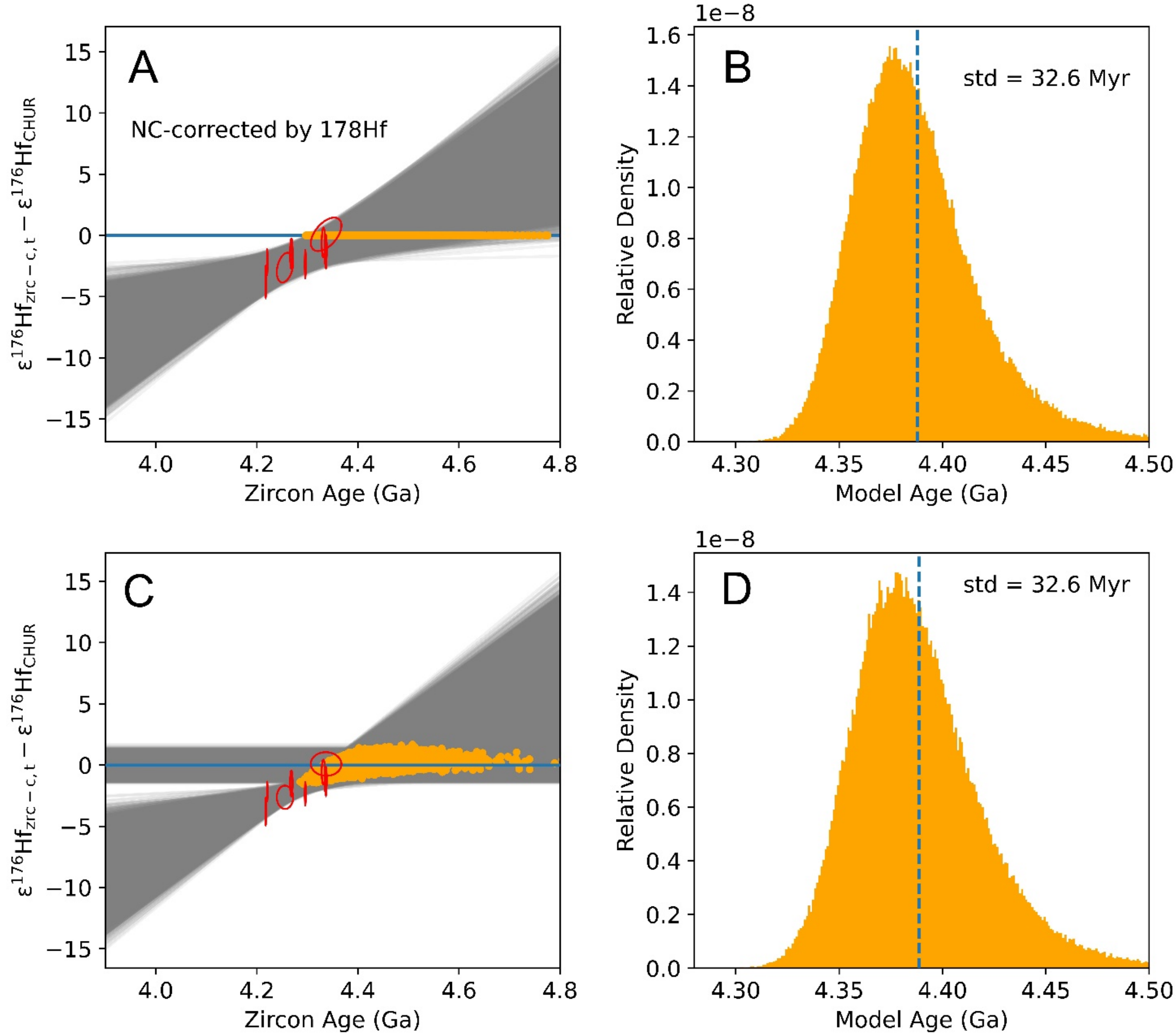

**Figure 9**. Example of how a zircon population model age can be calculated by regressing $\varepsilon^{176}$Hf against age (here neutron capture effects have been corrected for using $\varepsilon^{178}$Hf). Panel A shows an approach that includes uncertainties on CHUR parameters in the $\varepsilon^{176}$Hf value of each zircon, meaning that errors affecting different zircons will not be independent (Eq. 18). Panel C shows an approach that separates errors affecting zircons from CHUR (Eqs. 22 and 23). A Monte Carlo approach was used to calculate the reservoir model age and its uncertainty by calculating the intercept between a regression through the data and the CHUR reference. The histogram of the model ages from intercept points are plotted in panel B and D.

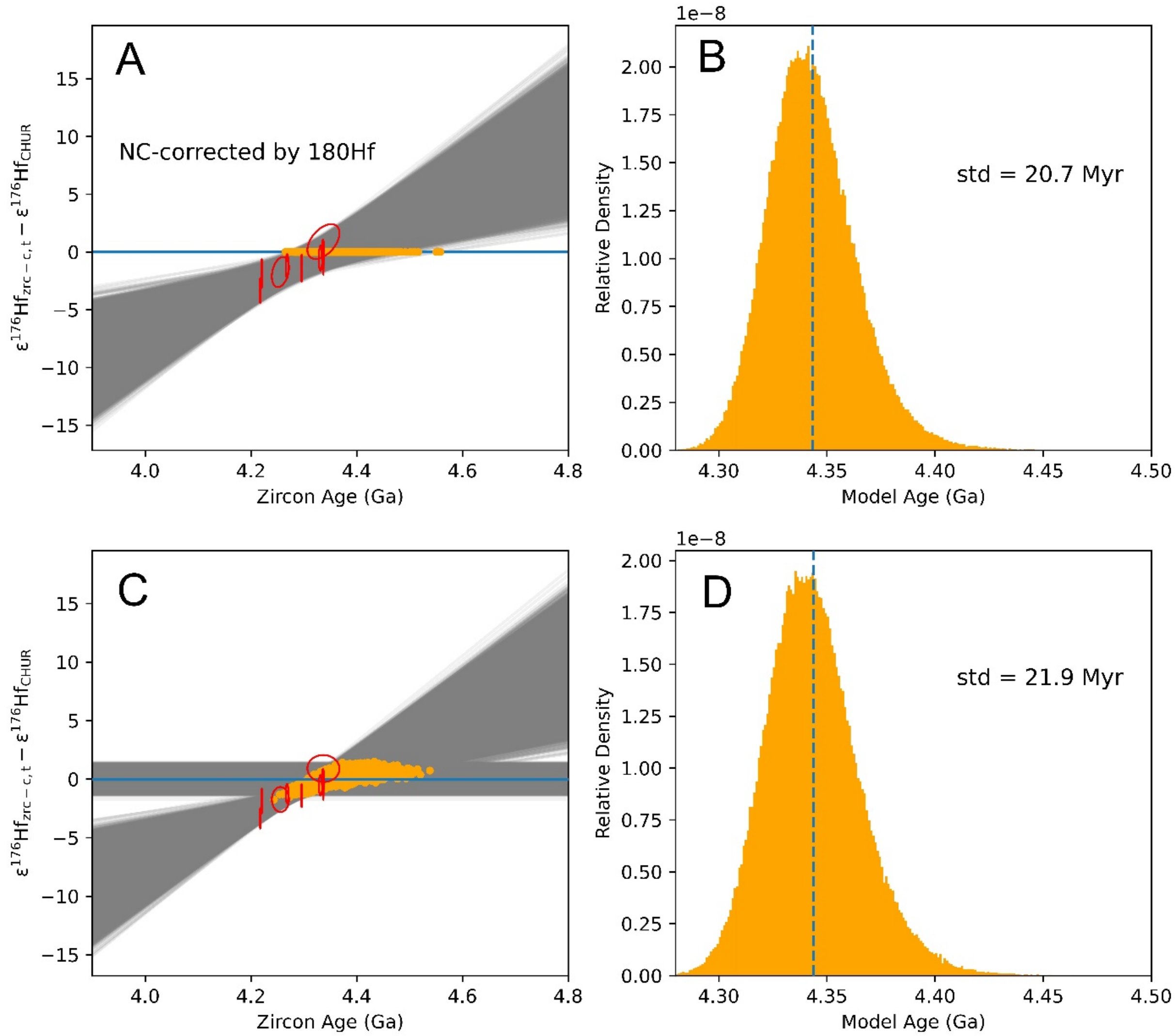

**Figure 10**. Example of how a zircon population model age can be calculated by regressing $\varepsilon^{176}$Hf against age (here neutron capture effects have been corrected for using $\varepsilon^{180}$Hf). Panel A shows an approach that includes uncertainties on CHUR parameters in the $\varepsilon^{176}$Hf value of each zircon, meaning that errors affecting different zircons will not be independent (Eq. 18). Panel C shows an approach that separates errors affecting zircons from CHUR (Eqs. 22 and 23). A Monte Carlo approach was used to calculate the reservoir model age and its uncertainty by calculating the intercept between a regression through the data and the CHUR reference. The histogram of the model ages from intercept points are plotted in panel B and D.

**Table 1.**

**Chromatographic purification protocol for Zr and Hf**

| **Step** | **Volume (mL)** | **Acid** | **Step** | **Volume (mL)** | **Acid** |
|---|---|---|---|---|---|
| **column 1** (2 mL TODGA; 0.8 cm diameter × 4 cm length) | | | **column 2** (0.35 mL Ln-Spec; 1.5 mm diameter × 20 cm length) | | |
| Clean | 10 | 3 M $HNO_3$ | Clean | 18 | 6 M HCl-0.06 M HF |
| | 10 | 3 M $HNO_3$-1 vol% $H_2O_2$ | | 14 | 6 M HCl-0.2 M HF |
| | 4 | $H_2O$ | Precondition | 6 | 2.5 M HCl |
| Precondition | 15 | 12 M $HNO_3$ | Load | 0.5 | 2.5 M HCl |
| Load | 10 | 12 M $HNO_3$ | Rinse matrix | 12 | 6 M HCl-1 vol% $H_2O_2$ |
| Rinse matrix | 10 | 12 M $HNO_3$ | Elute Zr | 22 | 6 M HCl-0.06 M HF |
| Elute Ti | 10 | 12 M $HNO_3$-1 vol% $H_2O_2$ | Hf | 7 | 6 M HCl-0.2 M HF |
| Fe | 10 | 3 M $HNO_3$ | | | |
| Zr and Hf | 20 | 3 M $HNO_3$-0.3M HF | | | |

**Table 2. The zircon reference materials analyzed in this study**

| Sample | Hf (ng) | $^{176}Hf/^{177}Hf$ | ±2$\sigma$ | $^{178}Hf/^{177}Hf$ | ±2$\sigma$ | $\Delta\varepsilon^{176}Hf$[a] | ±2$\sigma$ |
|---|---|---|---|---|---|---|---|
| AS3 | 3 | 0.282206 | 14 | 1.467166 | 26 | 0.80 | 0.48 |
| AS3 | 3 | 0.282221 | 13 | 1.467186 | 39 | 1.30 | 0.45 |
| AS3 | 10 | 0.282193 | 29 | 1.467173 | 25 | 0.32 | 1.04 |
| AS3 | 10 | 0.282194 | 10 | 1.467175 | 19 | 0.36 | 0.37 |
| AS3 | 8 | 0.282203 | 14 | 1.467166 | 14 | 0.67 | 0.51 |
| AS3 | 8 | 0.282196 | 8 | 1.467174 | 20 | 0.43 | 0.28 |
| AS3 | 18 | 0.282200 | 43 | 1.467188 | 133 | 0.55 | 1.53 |
| AS3 | 18 | 0.282197 | 5 | 1.467170 | 15 | 0.46 | 0.19 |
| AS3 | 20 | 0.282196 | 6 | 1.467181 | 12 | 0.44 | 0.21 |
| AS3 | 28 | 0.282198 | 4 | 1.467179 | 11 | 0.49 | 0.14 |
| **mean AS3** | | **0.282200** | **5** | **1.467176** | **5** | **0.58** | **0.17** |
| **FC-1**[b] | | **0.282184** | **8** | | | | |
| 91500 | 3 | 0.282317 | 20 | 1.467175 | 55 | 0.38 | 0.72 |
| 91500 | 3 | 0.282327 | 20 | 1.467174 | 27 | 0.74 | 0.70 |
| 91500 | 10 | 0.282312 | 8 | 1.467177 | 23 | 0.22 | 0.29 |
| | 10 | 0.282308 | 13 | 1.467183 | 21 | 0.08 | 0.45 |
| 91500 | 8 | 0.282304 | 5 | 1.467179 | 16 | -0.08 | 0.18 |
| 91500 | 18 | 0.282310 | 4 | 1.467176 | 16 | 0.14 | 0.15 |
| 91500 | 20 | 0.282308 | 4 | 1.467174 | 14 | 0.09 | 0.13 |
| 91500 | 28 | 0.282306 | 4 | 1.467179 | 8 | 0.01 | 0.13 |
| **mean 91500** | | **0.282312** | **5** | **1.467177** | **2** | **0.20** | **0.17** |
| **91500** | | **0.282306** | **4** | | | | |
| NZ32a | 3 | 0.282157 | 19 | 1.467162 | 50 | 0.59 | 0.68 |
| NZ32a | 3 | 0.282143 | 29 | 1.467171 | 25 | 0.10 | 1.03 |
| NZ32a | 10 | 0.282128 | 36 | 1.467175 | 37 | -0.42 | 1.29 |
| NZ32a | 8 | 0.282131 | 9 | 1.467158 | 11 | -0.33 | 0.33 |
| NZ32a | 18 | 0.282137 | 5 | 1.467167 | 12 | -0.11 | 0.19 |
| NZ32a | 20 | 0.282134 | 4 | 1.467162 | 11 | -0.22 | 0.14 |
| NZ32a | 28 | 0.282135 | 3 | 1.467166 | 8 | -0.17 | 0.12 |
| **mean NZ32a** | | **0.282138** | **7** | **1.467166** | **4** | **-0.08** | **0.24** |
| **NZ32a** | | **0.282140** | **5** | **1.467295** | **15** | | |

[a]$\Delta\varepsilon^{176}Hf = [(^{176}Hf/^{177}Hf_{measured})/(^{176}Hf/^{177}Hf_{literature}) - 1] * 10^4$
[b]AS3 is from the same geological unit as FC-1

**Table 3. Hf isotopic compositions and $^{207}$Pb-$^{206}$Pb ages of five lunar zircons (L1, L2 refer to the first and second leachate, while R refers to residues)**

| samples | $^{176}$Hf/$^{177}$Hf | 2σ* | $^{178}$Hf/$^{177}$Hf | 2σ* | $^{180}$Hf/$^{177}$Hf | 2σ* | $^{176}$Lu/$^{177}$Hf | 2σ | t Ma | 2σ |
|---|---|---|---|---|---|---|---|---|---|---|
| 14163 Z89 R | 0.280077 | 0.000023 | 1.467406 | 0.000034 | 1.886312 | 0.000110 | 0.000822 | 0.000028 | 4295.9 | 0.8 |
| 14163 Z9_L1 | 0.280117 | 0.000023 | 1.467192 | 0.000034 | 1.886651 | 0.000110 | 0.001958 | 0.000071 | 4268.3 | 2.4 |
| 14163 Z26_L1 | 0.280062 | 0.000023 | 1.467316 | 0.000034 | 1.886476 | 0.000110 | 0.000648 | 0.000022 | 4337.1 | 30.3 |
| 14163 Z26_L2 | 0.280037 | 0.000023 | 1.467314 | 0.000034 | 1.886489 | 0.000110 | 0.000619 | 0.000012 | 4255.7 | 16.2 |
| 14321 Z3_L1 | 0.280087 | 0.000009 | 1.467183 | 0.000014 | 1.886674 | 0.000098 | 0.001496 | 0.000240 | 4220.5 | 0.6 |
| 14321 Z3_L2 | 0.280092 | 0.000014 | 1.467185 | 0.000063 | 1.886687 | 0.000101 | 0.002074 | 0.000091 | 4217.5 | 0.5 |
| 72275 Z1_L1 | 0.280017 | 0.000014 | 1.467201 | 0.000063 | 1.886606 | 0.000101 | 0.000902 | 0.000041 | 4331.6 | 3.3 |
| 72275 Z1_L2 | 0.280008 | 0.000014 | 1.467234 | 0.000063 | 1.886678 | 0.000101 | 0.000868 | 0.000097 | 4336.2 | 2.1 |
| 72275 Z1 R | 0.280004 | 0.000023 | 1.467220 | 0.000034 | 1.886661 | 0.000110 | 0.001000 | 0.000026 | 4336.8 | 0.5 |

*errors reported here are based on the external reproducibilities of the JMC-475 Hf standard.

**Table 4. Comparisons between the single zircon neutron capture induced $\varepsilon^{176}Hf_{CHUR}$ shifts using different methods and models**

| | $\Delta\varepsilon^{176}Hf/^{177}Hf$ | | | |
|---|---|---|---|---|
| | Sprung theoretical model | | This study | |
| samples | based on $\varepsilon^{178}Hf/^{177}Hf$ | based on $\varepsilon^{180}Hf/^{177}Hf$ | based on $\varepsilon^{178}Hf/^{177}Hf$ | based on $\varepsilon^{180}Hf/^{177}Hf$ |
| 14163 Z89 R | -4.24 | -3.10 | -3.82 | -2.89 |
| 14163 Z9_L1 | -0.43 | -0.13 | -0.39 | -0.12 |
| 14163 Z26_L1 | -2.63 | -1.66 | -2.37 | -1.55 |
| 14163 Z26_L2 | -2.59 | -1.55 | -2.33 | -1.44 |
| 14321 Z3_L1 | -0.26 | 0.07 | -0.23 | 0.07 |
| 14321 Z3_L2 | -0.30 | 0.18 | -0.27 | 0.17 |
| 72275 Z1_L1 | -0.59 | -0.52 | -0.53 | -0.49 |
| 72275 Z1_L2 | -1.18 | 0.11 | -1.06 | 0.10 |
| 72275 Z1 R | -0.93 | -0.04 | -0.84 | -0.04 |

**Table 5. Calculated ε$^{176}$Hf and model ages of 5 lunar zircons.**

| Samples | ε$^{176}$Hf $_{Zrc_t_c/CHUR-t}$ | 2σ (Monte Carlo) | 2σ (Analytical Eqn.) | $\sigma_{f_1}^2$ | $\sigma_{f_4}^2$ | $\sigma_{f_2}^2$ | $\sigma_{f_3}^2$ | Minimum Model Age $t_d$ (Ma) $^{176}$Lu/$^{177}$Hf$_{_R_p}$ = 0 | 2σ (Ma) | Model Age $t_d$ (Ma) $^{176}$Lu/$^{177}$Hf$_{_R_p}$ = 0.0153±0.0033 | 2σ (Ma) |
|---|---|---|---|---|---|---|---|---|---|---|---|
| NC-178 correction | | | | | | | | | | | |
| 14163 Z89 R | -2.34 | 1.15 | 1.15 | 0.33 | 0.33 | 0.23 | 0.10 | 4391 | 47 | 4470 | 87 |
| 14163 Z9_L1 | -1.50 | 1.23 | 1.23 | 0.38 | 0.38 | 0.27 | 0.10 | 4330 | 50 | 4380 | 98 |
| 14163 Z26_L1 | 0.09 | 1.37 | 1.37 | 0.47 | 0.47 | 0.23 | 0.24 | 4334 | 47 | 4331 | 99 |
| 14163 Z26_L2 | -2.63 | 1.22 | 1.21 | 0.37 | 0.37 | 0.23 | 0.14 | 4363 | 47 | 4452 | 89 |
| 14321 Z3_L1 | -2.17 | 1.03 | 1.03 | 0.27 | 0.26 | 0.16 | 0.10 | 4309 | 42 | 4383 | 83 |
| 14321 Z3_L2 | -3.79 | 1.29 | 1.28 | 0.42 | 0.41 | 0.31 | 0.10 | 4373 | 53 | 4501 | 96 |
| 72275 Z1_L2 | -0.57 | 1.21 | 1.20 | 0.37 | 0.36 | 0.26 | 0.11 | 4355 | 49 | 4375 | 96 |
| 72275 Z1_L1 | -1.22 | 1.24 | 1.24 | 0.38 | 0.38 | 0.28 | 0.10 | 4386 | 51 | 4427 | 96 |
| 72275 Z1 R | -1.53 | 1.15 | 1.15 | 0.33 | 0.33 | 0.23 | 0.10 | 4400 | 47 | 4452 | 88 |
| NC-180 correction | | | | | | | | | | | |
| 14163 Z89 R | -1.41 | 1.16 | 1.16 | 0.34 | 0.34 | 0.23 | 0.10 | 4353 | 47 | 4401 | 91 |
| 14163 Z9_L1 | -1.23 | 1.04 | 1.05 | 0.27 | 0.27 | 0.17 | 0.10 | 4319 | 43 | 4361 | 86 |
| 14163 Z26_L1 | 0.91 | 1.51 | 1.51 | 0.57 | 0.57 | 0.33 | 0.24 | 4300 | 54 | 4269 | 115 |
| 14163 Z26_L2 | -1.74 | 1.30 | 1.30 | 0.42 | 0.42 | 0.28 | 0.14 | 4327 | 51 | 4386 | 99 |
| 14321 Z3_L1 | -1.87 | 1.22 | 1.22 | 0.37 | 0.37 | 0.27 | 0.10 | 4297 | 50 | 4360 | 98 |
| 14321 Z3_L2 | -3.35 | 1.05 | 1.06 | 0.28 | 0.28 | 0.17 | 0.10 | 4355 | 43 | 4468 | 81 |
| 72275 Z1_L2 | -0.53 | 1.08 | 1.08 | 0.29 | 0.29 | 0.19 | 0.11 | 4353 | 44 | 4371 | 88 |
| 72275 Z1_L1 | -0.06 | 1.12 | 1.12 | 0.31 | 0.31 | 0.21 | 0.10 | 4338 | 46 | 4340 | 93 |
| 72275 Z1 R | -0.73 | 1.20 | 1.20 | 0.36 | 0.36 | 0.26 | 0.10 | 4367 | 49 | 4392 | 95 |

**For TOC Only (Table of Contents Graphic)**

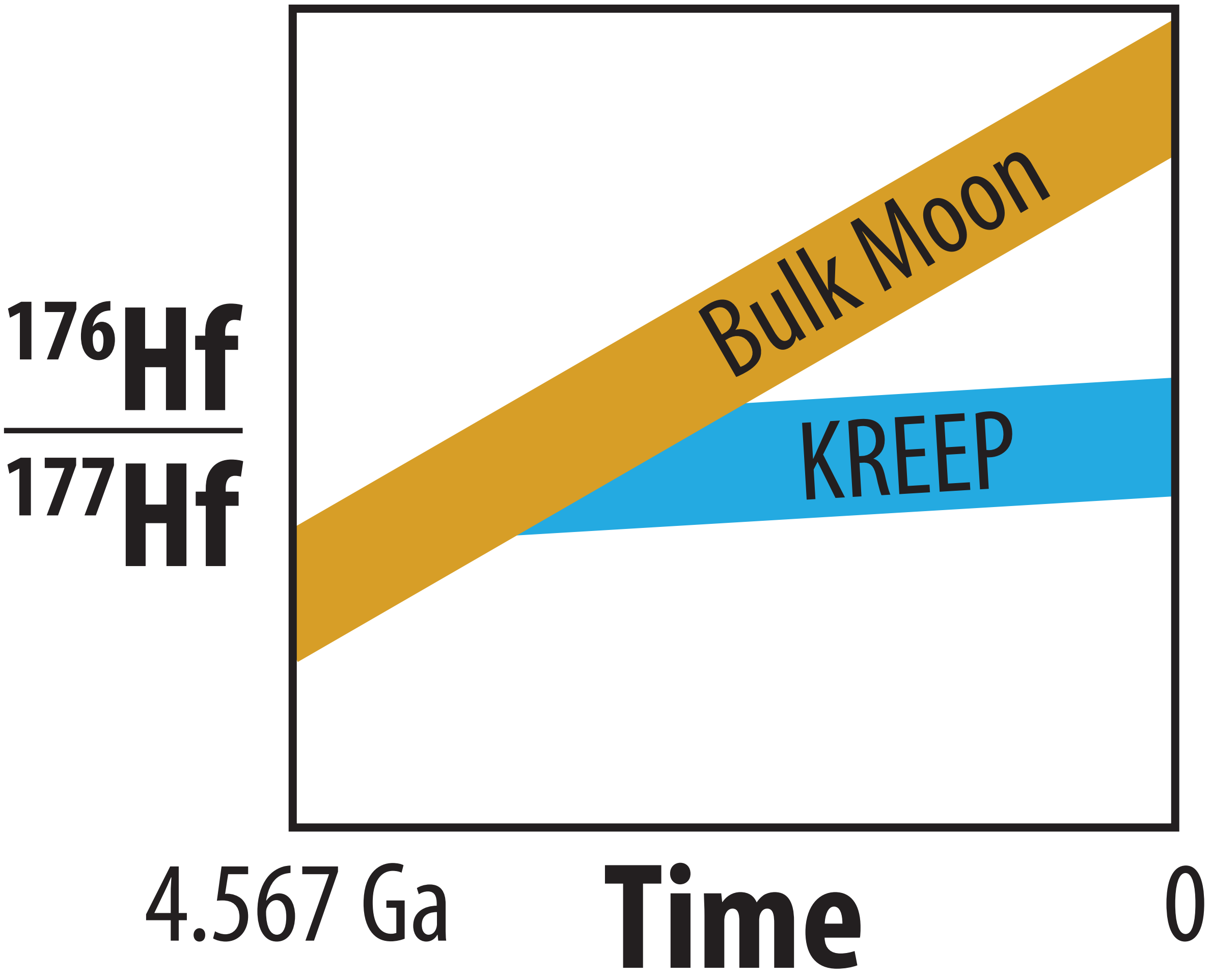

# Supporting Information

## Methodologies for $^{176}$Lu-$^{176}$Hf Analysis of Zircon Grains from the Moon and Beyond

Xi Chen[1], Nicolas Dauphas[1*], Zhe J. Zhang[1], Blair Schoene[2], Melanie Barboni[3], Ingo Leya[4], Junjun Zhang[1], Dawid Szymanowski[2], Kevin D. McKeegan[5]

[1]Origins Laboratory, Department of the Geophysical Sciences and Enrico Fermi Institute, The University of Chicago, Chicago, IL 60637, USA

[2]Department of Geosciences, Princeton University, Princeton, NJ 08544, USA

[3]CLAS-NS Departments, Arizona State University, Tempe, AZ 85281, USA

[4]Physics Institute, University of Bern, Sidlerstrasse 5, 3012 Bern, Switzerland

[5]Department of Earth, Planetary, and Space Sciences, University of California, Los Angeles, CA 90095, USA

**1. Initial Hf isotopic composition**

As discussed in the main text, we express $\varepsilon^{176}\mathrm{Hf}_{\mathrm{zrc}\text{-}t,c}$, $\varepsilon^{176}\mathrm{Hf}_{\mathrm{CHUR}\text{-}t}$, and the difference $\varepsilon^{176}\mathrm{Hf}_{\mathrm{zrc}\text{-}t,c} - \varepsilon^{176}\mathrm{Hf}_{\mathrm{CHUR}\text{-}t}$ as functions $(f)$ of several random variables $(x)$ and a constant $(C)$,

$$\varepsilon^{176}\mathrm{Hf}_{\mathrm{zrc}\text{-}t,c/\widetilde{\mathrm{CHUR}}\text{-t}} = f_2 = \left[\frac{x_1 - x_4(e^{x_5x_6}-1)}{C} - 1\right] \times 10^4 - \frac{x_1x_2x_3}{C}, \quad (22)$$

$$\varepsilon^{176}\mathrm{Hf}_{\mathrm{CHUR}\text{-}t/\widetilde{\mathrm{CHUR}}\text{-t}} = f_3 = \left[\frac{x_7 + x_8(e^{x_5x_9} - e^{x_5x_6})}{C} - 1\right] \times 10^4, \quad (23)$$

$$\varepsilon^{176}\mathrm{Hf}_{\mathrm{zrc}\text{-}t,c/\widetilde{\mathrm{CHUR}}\text{-t}} - \varepsilon^{176}\mathrm{Hf}_{\mathrm{CHUR}\text{-}t/\widetilde{\mathrm{CHUR}}\text{-t}} = f_4 = \frac{x_1 - x_4(e^{x_5x_6}-1) - x_7 - x_8(e^{x_5x_9} - e^{x_5x_6})}{C} \times 10^4 - \frac{x_1x_2x_3}{C}, \quad (24)$$

with $C$ calculated from the mean values of some variables,

$$C = \widetilde{x_7} + \widetilde{x_8}\left(e^{\widetilde{x_5}\widetilde{x_9}} - e^{\widetilde{x_5}\widetilde{x_6}}\right). \quad \text{(S1)}$$

The mean and standard deviation of each variable are given in the main text. Uncertainty propagation is complicated by the fact that some uncertainties are correlated, which can be tackled using the delta method[1].

1.1. Error propagation in function $f_2$

The formula that gives the variance of $f_2$ is,

$$\sigma_{f_2}^2 \simeq \nabla g_2 \times V_2 \times \nabla g_2^T, \quad \text{(S2)}$$

where $V_2$ is the covariance matrix,

$$V_2 = \begin{bmatrix} v(x_1,x_1) & 0 & v(x_1,x_3) & 0 & 0 & 0 \\ 0 & v(x_2,x_2) & 0 & 0 & 0 & 0 \\ v(x_1,x_3) & 0 & v(x_3,x_3) & 0 & 0 & 0 \\ 0 & 0 & 0 & v(x_4,x_4) & 0 & 0 \\ 0 & 0 & 0 & 0 & v(x_5,x_5) & 0 \\ 0 & 0 & 0 & 0 & 0 & v(x_6,x_6) \end{bmatrix}, \quad \text{(S3)}$$

and $\nabla g_2$ is the gradient vector,

$$\nabla g_2 = \begin{bmatrix} \frac{\partial f_2}{\partial x_1} & \cdots & \frac{\partial f_2}{\partial x_6} \end{bmatrix}. \quad \text{(S4)}$$

The only non-zero term in the covariance matrix is $v(x_1, x_3)$, as some dependence exists between $^{176}Hf/^{177}Hf$ and $^{178,180}Hf/^{177}Hf$ ratios through sharing of common isotopes in the internal normalization scheme. To calculate this correlation coefficient, we again use the delta method on the bracketing-standard normalization. The internally normalized $^{176}Hf/^{177}Hf$ ratio of a zircon ($S_1$) is normalized to the two measured ratios of bracketing standards $S_2$ and $S_3$, and their known absolute $^{176}Hf/^{177}Hf$ ratio $s$ through,

$$x_1 = \frac{2sS_1}{S_2+S_3}. \quad \text{(S5)}$$

$s$ is the $^{176}$Hf/$^{177}$Hf ratio of JMC-475 Hf of 0.282160. Similarly, the measured $\varepsilon^{i}$Hf values of zircons were calculated from internally normalized $\left({}^{i}\mathrm{Hf}/{}^{177}\mathrm{Hf}\right)_{\text{zrc-p}}$ ratios (R; $i$ = 178 or 180), which are also bracketed by two internally normalized bracketing standards ($R_2$ and $R_3$),

$$x_3 = 10^4\left(\frac{2R_1}{R_2+R_3} - 1\right). \tag{S6}$$

We note $g$ as the function that calculates $^{176}$Hf/$^{177}$Hf ($x_1$) and $\varepsilon^{178/180}$Hf ($x_3$) normalized by standard bracketing from the internally normalized ratio $(S_1, S_2, S_3, R_1, R_2, R_3)$,

$$g(S_1, S_2, S_3, R_1, R_2, R_3) = (x_1, x_3). \tag{S7}$$

The gradient matrix for $(x_1, x_3)$ can be written as,

$$\nabla g = \begin{bmatrix} \frac{2s}{S_2+S_3} & \frac{-2sS_1}{(S_2+S_3)^2} & \frac{-2sS_1}{(S_2+S_3)^2} & 0 & 0 & 0 \\ 0 & 0 & 0 & \frac{10^4}{(R_2+R_3)} & \frac{-2\times10^4 R_1}{(R_2+R_3)^2} & \frac{-2\times10^4 R_1}{(R_2+R_3)^2} \end{bmatrix}. \tag{S8}$$

The covariance matrix of $(S_1, S_2, S_3, R_1, R_2, R_3)$ takes the form,

$$\epsilon = \begin{bmatrix} cov(S_1,\mathrm{S}_1) & 0 & 0 & cov(S_1,R_1) & 0 & 0 \\ 0 & cov(S_2,\mathrm{S}_2) & 0 & 0 & cov(S_2,\mathrm{R}_2) & 0 \\ 0 & 0 & cov(S_3,S_3) & 0 & 0 & cov(S_3,\mathrm{R}_3) \\ cov(S_1,R_1) & 0 & 0 & cov(R_1,R_1) & 0 & 0 \\ 0 & cov(S_2,\mathrm{R}_2) & 0 & 0 & cov(R_2,R_2) & 0 \\ 0 & 0 & cov(S_3,\mathrm{R}_3) & 0 & 0 & cov(R_3,R_3) \end{bmatrix}.$$

(S9)

The null entries in this matrix stem from the fact that measurements made at different times are expected to be independent. The covariance matrix for $x_1$ and $x_3$ is,

$$v(x_1, x_3) = \begin{bmatrix} \sigma_1\sigma_1 & \rho\sigma_1\sigma_3 \\ \rho\sigma_1\sigma_3 & \sigma_3\sigma_3 \end{bmatrix}. \tag{S10}$$

According to the delta method, the covariance matrix can be approximated by,

$$v(x_1, x_3) = \begin{bmatrix} cov(x_1,x_1) & cov(x_1,x_3) \\ cov(x_1,x_3) & cov(x_3,x_3) \end{bmatrix} = \nabla g \cdot \epsilon \cdot \nabla g^{\mathrm{T}}, \tag{S11}$$

This can then be injected in Eq. S11 to S2 and we have,

$$\sigma_{f_2}^2 \simeq \left(\frac{10^4}{C}\right)^2 \left[\left(1 - \frac{x_2 x_3}{10^4}\right)^2 \sigma_{x_1}^2 + (e^{x_5 x_6} - 1)^2 \sigma_{x_4}^2 + (x_4 x_6 e^{x_5 x_6})^2 \sigma_{x_5}^2 + (x_4 x_5 e^{x_5 x_6})^2 \sigma_{x_6}^2\right] +$$

$$\frac{x_1^2 x_3^2}{c^2} \sigma_{x_2}^2 + \frac{x_1^2 x_2^2}{c^2} \sigma_{x_3}^2 + 2\frac{10^4}{C^2}\left(1 - \frac{x_2 x_3}{10^4}\right) x_1 x_3 \rho \sigma_{x_1 x_3}, \quad \text{(S12)}$$

The correlation coefficients ($\rho$) are calculated based on the measured $^{176}$Hf/$^{177}$Hf and $^{178/180}$Hf/$^{177}$Hf, and they are given in **Table S1**.

<u>1.2. Error propagation in function $f_3$</u>

Most of the parameters in $f_3$ are clearly independent. The solar system initial $(^{176}\text{Hf}/^{177}\text{Hf})_{\text{CHUR-ss}}$ ratio ($x_7$) from Iizuka et al. (2015) is calculated using Lu-Hf isotopes and Pb-Pb ages from eucrite zircons. The present $(^{176}\text{Lu}/^{177}\text{Hf})_{\text{CHUR-}p}$ was derived from chondrite measurements ($x_5$), while the solar system ages ($x_9$) were independently constrained by Pb-Pb dating[2]. The covariance matrix for function $f_3$ therefore takes the form,

$$V_3 = \begin{bmatrix} v(x_5, x_5) & \cdots & 0 \\ \vdots & \ddots & \vdots \\ 0 & \cdots & v(x_9, x_9) \end{bmatrix}. \quad \text{(S13)}$$

The gradient vector is,

$$\nabla g_3 = \begin{bmatrix} \frac{\partial f_3}{\partial x_5} & \cdots & \frac{\partial f_3}{\partial x_8} \end{bmatrix}. \quad \text{(S14)}$$

The uncertainty of $f_3$ is calculated using the delta method,

$$\sigma_{f_3}^2 \simeq \nabla g_3 \times V_3 \times \nabla g_3^T. \quad \text{(S15)}$$

and we have,

$$\sigma_{f_3}^2 \simeq \left(\frac{10^4}{C}\right)^2 \left[(x_8 x_9 e^{x_5 x_9} - x_8 x_6 e^{x_5 x_6})^2 \sigma_{x_5}^2 + (x_5 x_8 e^{x_5 x_6})^2 \sigma_{x_6}^2 + \sigma_{x_7}^2 + (e^{x_5 x_9} - e^{x_5 x_6})^2 \sigma_{x_8}^2 + (x_5 x_8 e^{x_5 x_9})^2 \sigma_{x_9}^2\right]. \quad \text{(S16)}$$

<u>1.3. Error propagation in function $f_4$</u>

As in 1.1, we have to deal with the fact that $x_1$ and $x_3$ are not independent. The steps used to calculate the uncertainty as in 1.1 and are not repeated here. The uncertainty on $\varepsilon^{176}\mathrm{Hf}_{\mathrm{zrc}\text{-}t,c} - \varepsilon^{176}\mathrm{Hf}_{\mathrm{CHUR}\text{-}t}$ ($f_4$) can be calculated as,

$$\sigma_{f_4}^2 \simeq \nabla g_4 \times V_4 \times \nabla g_4^T. \tag{S17}$$

with $V_4$ as the covariance matrix and $\nabla g_4$ as the gradient vectors. We therefore have,

$$\sigma_{f_4}^2 \simeq \left(\frac{10^4}{C}\right)^2 \left[\left(1-\frac{x_2x_3}{10^4}\right)^2 \sigma_{x_1}^2 + (e^{x_5x_6}-1)^2\sigma_{x_4}^2 + (x_4x_6e^{x_5x_6} + x_8x_9e^{x_5x_9} - x_8x_6e^{x_5x_6})^2\sigma_{x_5}^2 + (-x_4x_5e^{x_5x_6} + x_5x_8e^{x_5x_6})^2\sigma_{x_6}^2 + \sigma_{x_7}^2 + (e^{x_5x_9} - e^{x_5x_6})^2\sigma_{x_8}^2 + (x_5x_8e^{x_5x_9})^2\sigma_{x_9}^2\right] + \frac{x_1^2x_3^2}{c^2}\sigma_{x_2}^2 + \frac{x_1^2x_2^2}{c^2}\sigma_{x_3}^2 + 2\frac{10^4}{C^2}\left(1-\frac{x_2x_3}{10^4}\right)x_1x_3\sigma_{x_1x_3}. \tag{S18}$$

For each zircon, we have calculated the contribution of each variable $\left(\frac{\partial f_3}{\partial x_i}\right)^2 \sigma_{x_i}^2/\sigma_{f_3}^2$ and $2\frac{(\partial f_3)^2}{\partial x_1 \partial x_3}\sigma_{x_{1,3}}^2/\sigma_{f_3}^2$ to the overall variance of $f_4$ (**Table S1**). The main sources of error in $\varepsilon^{176}\mathrm{Hf}_{\mathrm{zrc}\text{-}t,c} - \varepsilon^{176}\mathrm{Hf}_{\mathrm{CHUR}\text{-}t}$ are the measured $(^{176}\mathrm{Hf}/^{177}\mathrm{Hf})_{\mathrm{zrc}\text{-}p}$ ratio ($x_1$), the measured isotopic shifts $\varepsilon^{178}\mathrm{Hf}$ and $\varepsilon^{180}\mathrm{Hf}$ that are used to correct cosmogenic effects ($x_3$), and the initial Hf isotopic composition of CHUR $(^{176}\mathrm{Hf}/^{177}\mathrm{Hf})_{\mathrm{CHUR}\text{-}ss}$ ($x_7$). Uncertainties from $x_2$,$x_4$,$x_5$,$x_6$,$x_8$, and $x_9$ are entirely negligible because they are either small or cancel out when we calculate the difference with CHUR. Neglecting those terms, we have a simpler expression for the error of $f_4$,

$$\sigma_{f_4}^2 \simeq \left(\frac{10^4}{C}\right)^2 \left[\left(1-\frac{x_2x_3}{10^4}\right)^2 \sigma_{x_1}^2 + \sigma_{x_7}^2\right] + \frac{x_1^2x_2^2}{c^2}\sigma_{x_3}^2 - 2\frac{10^4}{C^2}\left(1-\frac{x_2x_3}{10^4}\right)x_1x_3\sigma_{x_1x_3}. \tag{S19}$$

## 2. Model ages

In main text, we used eqn. 30 for the model ages for individual zircon, which in epsilon notation takes the form,

$$e^{\lambda_{^{176}\mathrm{Lu}} t_d} = e^{\lambda_{^{176}\mathrm{Lu}} t} + \frac{C}{10^4}\left[\frac{\varepsilon^{176}\mathrm{Hf}_{\text{zrc-t,c}} - \varepsilon^{176}\mathrm{Hf}_{\text{CHUR-}t}}{(^{176}\mathrm{Lu}/^{177}\mathrm{Hf})_{\text{R-}p} - (^{176}\mathrm{Lu}/^{177}\mathrm{Hf})_{\text{CHUR-}p}}\right]. \quad \text{(S20)}$$

Rearranging the Eq. S20, the model age can be written as:

$$t_d = t + \frac{1}{\lambda_{176_{Lu}}}\ln\left[\frac{\left(\frac{^{176}\mathrm{Hf}}{^{177}\mathrm{Hf}}\right)_{\text{zrc-p}} - \left(\frac{^{176}\mathrm{Lu}}{^{177}\mathrm{Hf}}\right)_{\text{zrc-}p}\left(e^{\lambda_{^{176}\mathrm{Lu}} t} - 1\right) - \left(\frac{^{176}\mathrm{Hf}}{^{177}\mathrm{Hf}}\right)_{\text{CHUR-}ss} - \left(\frac{^{176}\mathrm{Lu}}{^{177}\mathrm{Hf}}\right)_{\text{CHUR-}p}\left(e^{\lambda_{^{176}\mathrm{Lu}} t_{ss}} - e^{\lambda_{^{176}\mathrm{Lu}} t}\right) - \left(\frac{^{176}\mathrm{Hf}}{^{177}\mathrm{Hf}}\right)_{\text{zrc-p}}\left(\frac{\alpha_i \varepsilon^i\mathrm{Hf}}{10^4}\right)}{(^{176}\mathrm{Lu}/^{177}\mathrm{Hf})_{\text{R-}p} - (^{176}\mathrm{Lu}/^{177}\mathrm{Hf})_{\text{CHUR-}p}}\right]. \quad \text{(S21)}$$

Using the previously defined variables, we have,

$$t_d = \frac{1}{x_5}\ln\left[\frac{x_{10}e^{x_5x_6} + x_1 - x_4(e^{x_5x_6} - 1) - x_7 - x_8e^{x_5x_9} - \frac{x_1x_2x_3}{10^4}}{x_{10} - x_8}\right]. \quad \text{(S22)}$$

The standard deviation of $t_d$ can be calculated using the delta method,

$$\sigma_{t_d}^2 \simeq \sum\left(\frac{\partial f}{\partial x_i}\right)^2 \sigma_{x_i}^2 + 2\sum\sum\frac{(\partial f)^2}{\partial x_i \partial x_j}\sigma_{x_i x_j}. \quad \text{(S23)}$$

We now define $g = x_{10}e^{x_5x_6} + x_1 - x_4(e^{x_5x_6} - 1) - x_7 - x_8e^{x_5x_9} - \frac{x_1x_2x_3}{10^4}$, so Eq. S23 takes the form,

$$\sigma_{t_d}^2 \simeq \left[\frac{1}{gx_5}\left(1 - \frac{x_2x_3}{10^4}\right)\right]^2\sigma_{x_1}^2 + \left[\frac{1}{gx_5}\left(\frac{x_1x_3}{10^4}\right)\right]^2\sigma_{x_2}^2 + \left[\frac{1}{gx_5}\left(\frac{x_1x_2}{10^4}\right)\right]^2\sigma_{x_3}^2 + \left[\frac{1}{gx_5}(e^{x_5x_6} - 1)\right]^2\sigma_{x_4}^2 + \left[\frac{1}{x_5 g}(x_{10}x_6e^{x_5x_6} - x_4x_6e^{x_5x_6} - x_8x_9e^{x_5x_9}) - \frac{1}{x_5^2}\ln(gx_{10} - gx_8)\right]^2\sigma_{x_5}^2 + \left[\frac{1}{gx_5}(x_{10}x_6e^{x_5x_6} - x_4x_6e^{x_5x_6})\right]^2\sigma_{x_6}^2 + \left(\frac{1}{gx_5}\right)^2\sigma_{x_7}^2 + \left[\frac{1}{x_5}\left(\frac{1}{(x_{10}-x_8)} - \frac{e^{x_5x_9}}{g}\right)\right]^2\sigma_{x_8}^2 + \left(\frac{x_8x_9e^{x_5x_9}}{gx_5}\right)^2\sigma_{x_9}^2 + \left[\frac{1}{x_5}\left(\frac{(x_{10}-x_8)e^{x_5x_6}}{g} - \frac{1}{(x_{10}-x_8)}\right)\right]^2\sigma_{x_{10}}^2 - 2\times\left[\frac{1}{gx_5}\left(1 - \frac{x_2x_3}{10^4}\right)\right]\left[\frac{1}{gx_5}\left(\frac{x_1x_2}{10^4}\right)\right]\sigma_{x_1x_3}. \quad \text{(S24)}$$

We used this formula to calculate $\sigma_{t_d}^2$ and compared the results with Monte-Carlo simulations, and the two approaches agree (Figure S1). The main sources of errors for the model ages come from $x_1$, $x_3$, and $x_7$, which together contribute more than 99% to the total error (Table S1). The formulas are incorporated in Table S1.

**3. Model ages and initial Lu-Hf ratio**

The model age can also be calculated by doing a linear regression of $\varepsilon^{176}\mathrm{Hf}_{\text{zrc-}t,c}(t)$ versus the crystallization age $(t)$ of all or a subset of zircons. In this approach, the intersection between the zircon regression line and CHUR gives the model age and the slope reflects the $(^{176}\mathrm{Lu}/^{177}\mathrm{Hf})_{\text{R-}p}$ of the reservoir R. In the following text, we derive the analytical expression to the $(^{176}\mathrm{Lu}/^{177}\mathrm{Hf})_{\text{R-}p}$ and the model ages.

The $\varepsilon^{176}\mathrm{Hf}_{\text{zrc-t,c}}$ of zircons is a function of crystallization ages (t), and expressed as,

$$\varepsilon^{176}\mathrm{Hf}_{\text{zrc-t,c}} - \varepsilon^{176}\mathrm{Hf}_{\text{CHUR-}t} = \frac{10^4}{C}\left[\left(\frac{^{176}\mathrm{Lu}}{^{177}\mathrm{Hf}}\right)_{\text{R-}p} - \left(\frac{^{176}\mathrm{Lu}}{^{177}\mathrm{Hf}}\right)_{\text{CHUR-}p}\right]\left(e^{\lambda_{^{176}\mathrm{Lu}}t_d} - e^{\lambda_{^{176}\mathrm{Lu}}t}\right). \quad \text{(S24)}$$

The slope of regression line can be expressed as,

$$slope = \frac{10^4}{C}\left[\left(\frac{^{176}\mathrm{Lu}}{^{177}\mathrm{Hf}}\right)_{\text{R-}p} - \left(\frac{^{176}\mathrm{Lu}}{^{177}\mathrm{Hf}}\right)_{\text{CHUR-}p}\right]\frac{\left(e^{\lambda_{^{176}\mathrm{Lu}}t_d} - e^{\lambda_{^{176}\mathrm{Lu}}t}\right)}{t-t_d}. \quad \text{(S25)}$$

Rearranging the equation above, we obtain,

$$\left(\frac{^{176}\mathrm{Lu}}{^{177}\mathrm{Hf}}\right)_{\text{R-}p} = \left(\frac{^{176}\mathrm{Lu}}{^{177}\mathrm{Hf}}\right)_{\text{CHUR-}p} - \frac{C}{10^4}\frac{slope\times(t_d-t)}{\left(e^{\lambda_{^{176}\mathrm{Lu}}t_d} - e^{\lambda_{^{176}\mathrm{Lu}}t}\right)}. \quad \text{(S26)}$$

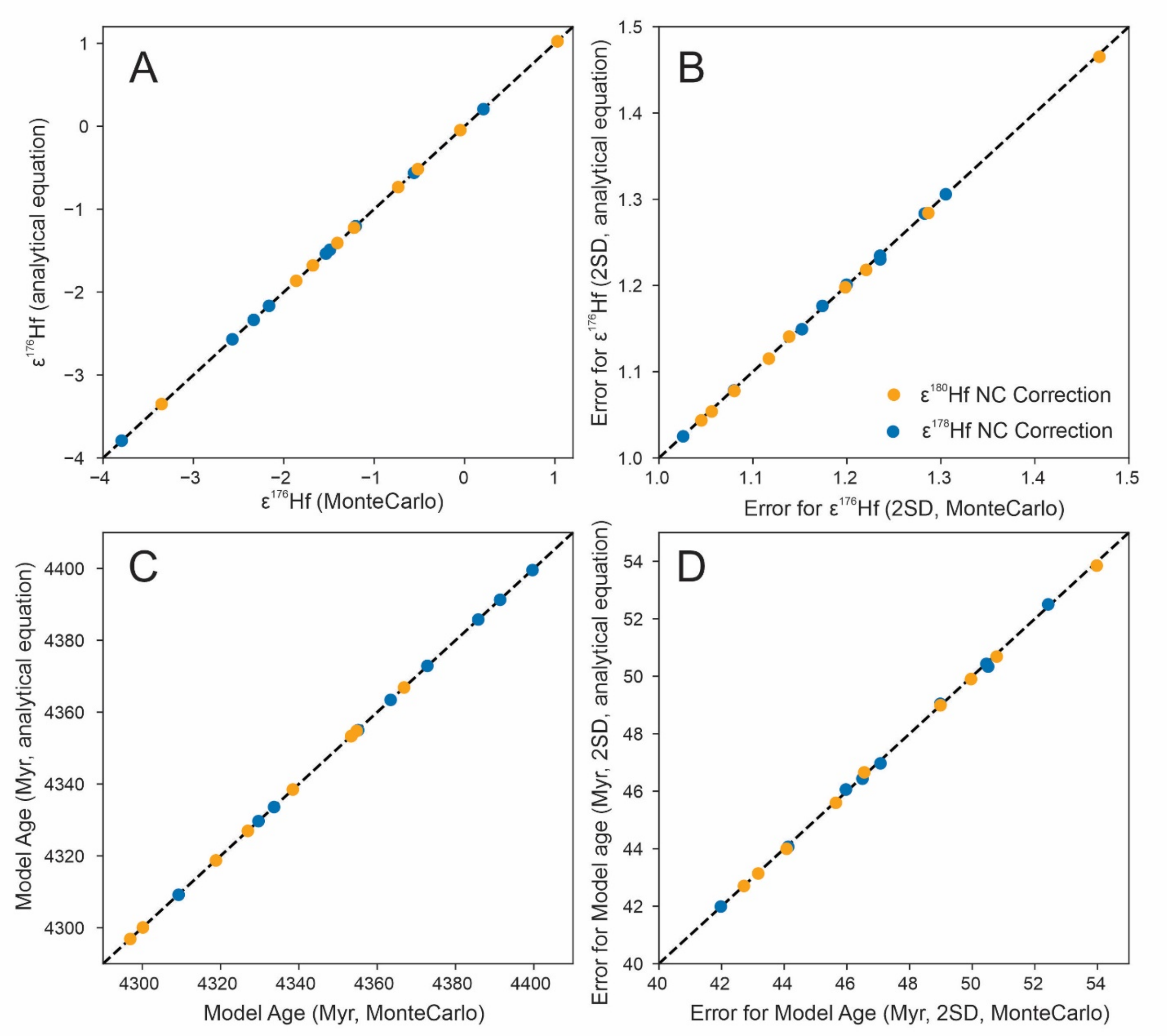

Figure S1. The comparison of calculated $\varepsilon^{176}$Hf (A), Model Ages (C) and their associated errors (B and D) between Monte Carlo simulation and the analytical equations. The two methods yield the same values.

## Table S1. Error contributions (%) of each term to the final error of $\varepsilon^{176}Hf_{CHUR-t}$ and model age of zircon data

| Sample | $\left(\frac{\partial f}{\partial x_1}\right)^2 \sigma^2 x_1/\sigma_f^2$ | | $\left(\frac{\partial f}{\partial x_2}\right)^2 \sigma^2 x_2/\sigma_f^2$ | | $\left(\frac{\partial f}{\partial x_3}\right)^2 \sigma^2 x_3/\sigma_f^2$ | | $\left(\frac{\partial f}{\partial x_4}\right)^2 \sigma^2 x_4/\sigma_f^2$ | | $\left(\frac{\partial f}{\partial x_5}\right)^2 \sigma^2 x_5/\sigma_f^2$ | | $\left(\frac{\partial f}{\partial x_6}\right)^2 \sigma^2 x_6/\sigma_f^2$ | | $\left(\frac{\partial f}{\partial x_7}\right)^2 \sigma^2 x_7/\sigma_f^2$ | | $\left(\frac{\partial f}{\partial x_8}\right)^2 \sigma^2 x_8/\sigma_f^2$ | | $\left(\frac{\partial f}{\partial x_9}\right)^2 \sigma^2 x_9/\sigma_f^2$ | | $2\frac{(\partial f_3)^2}{\partial x_1 \partial x_3}\sigma_{x_{1,3}}^2 /\sigma_{f_3}^2$ | |
|---|---|---|---|---|---|---|---|---|---|---|---|---|---|---|---|---|---|---|---|---|
| | **$\varepsilon^{176}$Hf** | $t_d$ | $\varepsilon^{176}$Hf | $t_d$ | $\varepsilon^{176}$Hf | $t_d$ | $\varepsilon^{176}$Hf | $t_d$ | $\varepsilon^{176}$Hf | $t_d$ | $\varepsilon^{176}$Hf | $t_d$ | $\varepsilon^{176}$Hf | $t_d$ | $\varepsilon^{176}$Hf | $t_d$ | $\varepsilon^{176}$Hf | $t_d$ | $\varepsilon^{176}$Hf | $t_d$ |
| NC-178 correction | | | | | | | | | | | | | | | | | | | | |
| 14163 Z89 | 57% | 57% | 1.1% | 1.1% | 25% | 25% | 0.6% | 0.6% | 0.0% | 0.1% | 0.0% | 0.0% | 36% | 36% | 0.0% | 0.0% | 0.0% | 0.0% | -19% | -19% |
| 14163 Z9_L1 | 43% | 43% | 0.0% | 0.0% | 19% | 19% | 2.9% | 2.9% | 0.0% | 0.1% | 0.2% | 0.0% | 27% | 27% | 0.0% | 0.0% | 0.0% | 0.0% | 7% | 7% |
| 14163 Z26_L1 | 39% | 51% | 0.3% | 0.4% | 17% | 23% | 0.3% | 0.3% | 0.0% | 0.1% | 24.5% | 0.0% | 24% | 32% | 0.0% | 0.0% | 0.0% | 0.0% | -5% | -7% |
| 14163 Z26_L2 | 47% | 52% | 0.3% | 0.4% | 21% | 23% | 0.1% | 0.1% | 0.0% | 0.1% | 8.3% | 0.0% | 30% | 33% | 0.0% | 0.0% | 0.0% | 0.0% | -7% | -8% |
| 14321 Z3_L1 | 10% | 10% | 0.0% | 0.0% | 5% | 5% | 47% | 47% | 0.0% | 0.2% | 0.0% | 0.0% | 39% | 39% | 0.1% | 0.0% | 0.0% | 0.0% | -1% | -1% |
| 14321 Z3_L2 | 16% | 16% | 0.0% | 0.0% | 63% | 63% | 4.3% | 4.3% | 0.0% | 0.1% | 0.0% | 0.0% | 25% | 25% | 0.0% | 0.0% | 0.0% | 0.0% | -8% | -8% |
| 72275 Z1_L1 | 18% | 18% | 0.0% | 0.0% | 72% | 72% | 1.1% | 1.1% | 0.0% | 0.1% | 0.3% | 0.0% | 29% | 29% | 0.0% | 0.0% | 0.0% | 0.0% | -20% | -20% |
| 72275 Z1_L2 | 17% | 17% | 0.1% | 0.1% | 68% | 68% | 5.6% | 5.6% | 0.0% | 0.1% | 0.1% | 0.0% | 27% | 27% | 0.0% | 0.0% | 0.0% | 0.0% | -18% | -18% |
| 72275 Z1 | 50% | 50% | 0.0% | 0.0% | 22% | 22% | 0.5% | 0.5% | 0.0% | 0.1% | 0.0% | 0.0% | 31% | 31% | 0.0% | 0.0% | 0.0% | 0.0% | -4% | -4% |
| NC-180 Correction | | | | | | | | | | | | | | | | | | | | |
| 14163 Z89 | 50% | 50% | 0.1% | 0.1% | 63% | 63% | 1% | 1% | 0.0% | 0.1% | 0.0% | 0.0% | 32% | 32% | 0.0% | 0.0% | 0.0% | 0.0% | -45% | -45% |
| 14163 Z9_L1 | 60% | 60% | 0.0% | 0.0% | 75% | 75% | 4% | 4% | 0.0% | 0.2% | 0.2% | 0.0% | 38% | 38% | 0.0% | 0.0% | 0.0% | 0.0% | -77% | -77% |
| 14163 Z26_L1 | 31% | 38% | 0.0% | 0.0% | 38% | 47% | 0% | 0% | 0.0% | 0.1% | 19.5% | 0.0% | 19% | 24% | 0.0% | 0.0% | 0.0% | 0.0% | -7% | -9% |
| 14163 Z26_L2 | 40% | 43% | 0.0% | 0.0% | 49% | 53% | 0% | 0% | 0.0% | 0.1% | 7.0% | 0.0% | 25% | 27% | 0.0% | 0.0% | 0.0% | 0.0% | -21% | -23% |
| 14321 Z3_L1 | 7% | 7% | 0.0% | 0.0% | 43% | 43% | 33% | 33% | 0.0% | 0.1% | 0.0% | 0.0% | 28% | 28% | 0.0% | 0.0% | 0.0% | 0.0% | -12% | -12% |
| 14321 Z3_L2 | 24% | 24% | 0.0% | 0.0% | 61% | 61% | 6% | 6% | 0.0% | 0.2% | 0.0% | 0.0% | 37% | 37% | 0.1% | 0.0% | 0.0% | 0.0% | -28% | -28% |
| 72275 Z1_L1 | 23% | 23% | 0.0% | 0.0% | 58% | 58% | 1% | 1% | 0.0% | 0.1% | 0.4% | 0.0% | 36% | 36% | 0.0% | 0.0% | 0.0% | 0.0% | -18% | -18% |
| 72275 Z1_L2 | 21% | 21% | 0.0% | 0.0% | 54% | 54% | 7% | 7% | 0.0% | 0.1% | 0.2% | 0.0% | 33% | 33% | 0.0% | 0.0% | 0.0% | 0.0% | -16% | -16% |
| 72275 Z1 | 46% | 46% | 0.0% | 0.0% | 57% | 57% | 0% | 0% | 0.0% | 0.1% | 0.0% | 0.0% | 29% | 29% | 0.0% | 0.0% | 0.0% | 0.0% | -32% | -32% |